\documentclass[manuscript, nonacm]{acmart}

\usepackage{array}
\usepackage{threeparttable}  
\usepackage{booktabs}
\usepackage{graphicx}
\usepackage{longtable}
\usepackage{placeins}

\AtBeginDocument{%
  \providecommand\BibTeX{{%
    \normalfont B\kern-0.5em{\scshape i\kern-0.25em b}\kern-0.8em\TeX}}}

\setcopyright{none}
\renewcommand\footnotetextcopyrightpermission[1]{} 

\begin{document}

\title[From Voices to Validity]{From Voices to Validity: Leveraging Large Language Models (LLMs) for Textual Analysis of Policy Stakeholder Interviews}

\author{Alex Liu}
\email{alexliux@uw.edu}
\orcid{0000-0002-4785-1801}
\author{Min Sun}
\authornotemark[1]
\email{misun@uw.edu}
\orcid{0000-0001-5832-1534}
\affiliation{%
  \institution{University of Washington}
  \streetaddress{2012 Skagit Lane, Miller Hall}
  \city{Seattle}
  \state{Washington}
  \country{USA}
  \postcode{98195-3600}
}

\begin{abstract}
 Obtaining stakeholders' diverse experiences and opinions about current policy in a timely manner is crucial for policymakers to identify strengths and gaps in resource allocation, thereby supporting effective policy design and implementation. However, manually coding even moderately sized interview texts or open-ended survey responses from stakeholders can often be labor-intensive and time-consuming. This study explores the integration of Large Language Models (LLMs)---like GPT-4---with human expertise to enhance text analysis of stakeholder interviews regarding K-12 education policy within one U.S. state. Employing a mixed-methods approach, human experts developed a codebook and coding processes as informed by domain knowledge and unsupervised topic modeling results. They then designed prompts to guide GPT-4 analysis and iteratively evaluate different prompts' performances. This combined human-computer method enabled nuanced thematic and sentiment analysis. Results reveal that while GPT-4 thematic coding aligned with human coding by 77.89\% at specific themes, expanding to broader themes increased congruence to 96.02\%, surpassing traditional Natural Language Processing (NLP) methods by over 25\%. Additionally, GPT-4 is more closely matched to expert sentiment analysis than lexicon-based methods. Findings from quantitative measures and qualitative reviews underscore the complementary roles of human domain expertise and automated analysis as LLMs offer new perspectives and coding consistency. The human-computer interactive approach enhances efficiency, validity, and interpretability of educational policy research.
\end{abstract}

\ccsdesc[500]{Human-centered computing~Human computer interaction (HCI)}
\ccsdesc[500]{Computing methodologies~Natural language processing}
\ccsdesc[500]{General and reference~Cross-computing tools and techniques}
\ccsdesc[500]{Social and professional topics~K-12 education}

\keywords{Education policy, Large Language Models (LLMs), Thematic analysis, Sentiment analysis, Policy stakeholders}

\received{02 December 2023}

\maketitle

\section{Introduction}
Policymakers seek reliable, valid, and meaningful evidence to support decision-making in a timely manner. An important source of policy evidence comes from stakeholders’ lived experiences regarding the implementation of current policies and their suggestions for improvement \cite{davidson2022improving, fedorowicz2021improving}. These stakeholders include individuals and organizations concerned with or affected by a policy’s creation, enactment, and evaluation. Insights can be gleaned from stakeholder interviews, open-ended surveys, or social media posts \cite{berry2013tensions, rosenberg2021understanding, wallner2008legitimacy}. Qualitative data on stakeholders’ insights, combined with quantitative causal analysis of policy impact, are often utilized in policy analysis. However, when decisions must be made swiftly, the cost of manually analyzing even a moderately sized corpus of text can impede the actual incorporation of stakeholders’ voices \cite{grimmer2013text}.

The rapid advancement of natural language processing (NLP) techniques---from traditional rule-based and statistical models to deep learning-based models, such as large language models (LLMs)---sheds light on the time-efficient comprehension of contents and sentiments in large text corpora. Traditional NLP approaches, such as topic modeling and specifically methods like Latent Dirichlet Allocation (LDA), have found favor in social sciences and policy studies for uncovering latent themes in extensive collections of texts, such as political speeches, news articles, and social media content \cite{blei2003latent, dimaggio2013exploiting, gao2023collabcoder, zhou2023guided}. In addition to content understanding, policy researchers are also interested in gauging stakeholders’ satisfaction about a given policy, perceived intended and unintended consequences, and areas for further improvement. Lexical-based sentiment analysis has been a popular tool in their analytical toolkit for exploring public reactions expressed on social media or other platforms \cite{abdulaziz2021topic, das2021comparative}. However, despite being easy to interpret, both LDA and lexical-based sentiment analysis encounter challenges in accuracy and adaptability when applied to domain-specific corpora.

Recent advances in LLMs have demonstrated impressive abilities to capture textual nuances. Researchers across various fields are examining the performance of LLMs in content and sentiment analysis tasks (e.g., \cite{chew2023llm, wang2023chatgpt}). Yet, few studies have focused on these promising tools within the context of educational policy analysis. As a recent and evolving technology, GPT-4’s performance varies across different domains and tasks. Thus, it is crucial for education policy researchers to explore how to adapt it for their domain-specific needs to fully benefit from it. Furthermore, given the “black box” nature of many proprietary deep learning-based models (such as ChatGPT), researchers must remain cautious of their limitations when employing these tools to make high-stake decisions in the public policy arena \cite{shlezinger2023model, tang2023science}. Before deploying GPT in policy analysis, it is incumbent upon us to investigate: How much can policymakers and researchers trust on large language models to analyze stakeholder feedback and opinions in high-stake, context-dependent decision making?

This paper is part of a broader study that identifies policies and programs advancing or hindering educational equity in Washington State (WA)'s K-12 public school system in 2022. In this study, educational equity is defined as the reduction of disparities in learning opportunities and outcomes for students from racial-ethnic minority groups and those of low income. To collect stakeholders’ lived experience and their opinions on this issue, the project conducted interviews with a diverse group of stakeholders, including state legislators, state-level policymakers, school district administrators, teacher union representatives, teachers, policy advocates, and community leaders. Our study, guided by two sets of research questions, seeks to understand and enhance the application and performance of computer-assisted textual analysis techniques in educational policy studies:\\
{\bfseries Substance Research Questions}
\begin{enumerate}
  \item What are the key themes that WA stakeholders voiced about the K-12 public school system?
  \item Which themes did stakeholders recognize as advancing educational equity (positive)? Conversely, which areas were mentioned as needing improvement or hinder (negative) educational equity?
\end{enumerate}
{\bfseries Methodological Research Questions}
\begin{enumerate}
  \item How accurate and valid are GPT-4 labels of key themes when comparing to human experts’ labels and traditional topic modeling results?
  \item How accurate and valid are GPT-4 sentiment classifications when comparing to human experts’ and lexicon-based sentiment analysis?
\end{enumerate}

In this study, we employ a multi-faceted analytical approach, integrating thematic and sentiment analysis executed by both human coders and NLP approaches, including GPT-4, LDA, and lexical-based methods. For Substance Research Question 1 (SRQ 1), we conceptualize a Resource Equity framework, which informs our expert-derived codebook for deductive thematic coding. The analysis captures stakeholder conversations around pivotal topics such as data accessibility, governance, anti-racism, and diversity. Our findings also reveal that the prominence of these themes is closely tied to stakeholders’ job roles, reflecting their direct interactions and concerns within the educational field.

For Substance Research Question 2 (SRQ 2), we delve into the sentiment landscape by categorizing responses into positive, neutral, and negative sentiments based on stakeholders’ satisfaction levels concerning educational policies and practices. The sentiment analysis highlights the demand for improvement in accountability, data access, and resource allocation. However, it also underscores a recognition and appreciation for the state’s recent advancements in multilingual education and student support systems, particularly among educators.

In addressing Methodological Research Question 1 (MRQ 1), we assess the efficacy of GPT-4 and LDA in thematic coding against human-coding, utilizing various metrics for evaluation. Our analysis demonstrates that GPT-4 not only generally surpasses LDA but also aligns with 77\% of the themes identified by human experts. Despite a tendency for GPT-4 to focus intently on certain codes and occasionally struggle with overlapping themes, our qualitative reviews of GPT-4’s results indicate it offers novel insights absent from human analysis, suggesting its potential as a valuable adjunct to human expertise.

For Methodological Research Question 2 (MRQ 2), we explore sentiment analysis through GPT-4, informed by domain-specific prompt design and benchmarked against human analysis, with the VEDAR lexical-based approach serving as a comparative tool. While both machine-led analyses exhibit challenges in fully grasping nuanced expressions of dissatisfaction in stakeholder narratives, GPT-4 successfully mirrors the majority of human-coded sentiments and exhibits the potential for bringing a fresh perspective for human coders. 

In the following sections, we will discuss previous studies that employed computer-assisted methods approaches for semantic discovery and coding. Then, we will outline the conceptual framework that informed the formation of our codebook, warranting that the coding process yields results relevant to the policy issue of interest. After describing the data collection and analysis process, we summarize our findings and discuss both policy and methodological implications. 

\section{Related Work}
Automated content analysis methods have made it possible to discover latent themes and understand underlying sentiments in stakeholders’ narratives about given policies by systematically analyzing large text collections without massive labor or time investment. Yet, the complexity of human language and domain-specific contexts suggest that automated content analysis cannot simply replace the nuanced and close reading provided by humans. Without appropriate guidance and supervision, the outputs of automated text analysis may be incomplete or misleading. Therefore, as the current technology stands, automatic methods should be thought of as amplifying and supplementing careful human analysis \cite{grimmer2013text}. Consequently, it is incumbent upon researchers to validate their use of automated text analysis. In this section, we review related prior work in GPT, traditional NLP approaches in textual analysis, their performance and limitations, and the unique contributions of our work.

\subsection{GPT and Textual Data Analysis}
With the advent of LLMs, researchers are rapidly exploring their potential and pitfalls for various tasks. Several studies have focused on thematic or sentiment analysis of text data using GPT models. By identifying and labeling latent themes and sentiments in text, GPT can potentially imitate or assist in qualitative analysis, which encompasses both inductive coding—building theoretical meaning from the text—and deductive coding—applying theoretical concepts to interpret text \cite{azungah2018qualitative}.

Gao et al. [2023]\cite{gao2023collabcoder} have incorporated GPT-3.5 in the inductive qualitative coding process to inform initial codebook formation and decision-making, demonstrating the potential for human-AI collaboration within the context of qualitative analysis. Dai et al. [2023]\cite{dai2023llm} extended this human-AI collaboration to complete the deductive coding process, utilizing a GPT-informed codebook to code survey responses about music tracks and password management. They evaluated the machine’s independent coding performance against human coding results, where both machine and human coders contributed to the codebook, and reported cosine similarities of 0.89, an average recall of 0.745, and Cohen’s Kappa (Cohen’s $\kappa$) above 0.8 for both datasets. Chew et al. [2023]\cite{chew2023llm} also indicated the feasibility of using GPT-3.5 for assisting deductive coding. Their study employed 4 publicly available data sets, comprising informal text like social media posts and formal text like news articles and achieved inter-rater reliability with Gwet’s AC1 ranging from 0.23 to 1.

Sentiment analysis has been performed using GPT with less labor input. Studies of social media posts and website reviews have shown that GPT-3.5 is capable of understanding sentiment and capturing tones, including sarcasm \cite{kheiri2023sentimentgpt, belal2023leveraging}. Belal et al.’s [2023]\cite{belal2023leveraging} study demonstrated that GPT-3.5 significantly enhanced the accuracy of three-category sentiment labeling for soccer tweets, increasing it from 47\% to 67\%, compared to the widely used lexical-based sentiment analysis tool, VADER.

However, GPT models tend to focus on certain aspects of the input text, resulting in errors and biases that can be problematic if these biases and errors systematically correlate with people’s characteristics, such as gender, education, race and ethnicity, and socioeconomic status \cite{ashwin2023using}. If bias is introduced during the human-AI collaboration of creating the codebook, it could jeopardize the validity of subsequent analyses using that codebook. In Dai et al.’s research that utilized a codebook developed through a collaboration between human and GPT, Cohen’s $\kappa$ lowered to 0.47 between GPT and a human coder who was not involved in the codebook’s creation, compared to a value of 0.81 between GPT with a human coder who was involved. This congruence diminished further to 0.34 when applying the codebook to data previously unseen by GPT during its training phase. Adaptation of expert-drafted codebooks can mitigate LLM-generated bias from the inductive coding process and prevent biases in codebooks from affecting the deductive coding step. Xiao et al. [2023]\cite{xiao2023supporting} combined an expert-developed codebook with GPT-3 coding to analyze children’s curiosity-driven questions and suggested the viability of this workflow. They reported Cohen’s $\kappa$ of 0.61 for labeling question complexity and 0.38 for syntactic structure. When comparing interrater reliability between linguistic experts, the LLM performance showed room for improvement in identifying linguistic concepts (Question Complexity: Cohen’s $\kappa$ = 0.88; Syntactic Structure: Cohen’s $\kappa$ = 0.90).

Furthermore, LLMs like GPT may not possess all domain-specific knowledge, and their performance also varies based on the structure and size of the codebook and the type of code. Chew et al. [2023]\cite{chew2023llm} found that when testing GPT-3.5 for deductive coding, paragraph-length blog posts coded using the most extensive set of codes (28 codes) in a hierarchical code set derived from concepts had the lowest inter-rater reliability (Gwet’s AC1 = 0.59), compared to news articles coded with 5 codes (Gwet’s AC1 = 0.85) and water quality reports with 5 codes of technically complex classes (average Gwet’s AC1 = 0.76). A recent study on GPT-3.5’s semantic annotation of legal texts found lower F1 scores \footnote{0.54 for the purpose of the public-health system’s emergency response and preparedness statutory and regulatory provisions, compared to 0.86 the types of contractual clauses.} associated with datasets requiring more domain expertise to discern nuances in the text and distinctions between codes \cite{savelka2023unlocking}. These weaknesses present risks for deploying GPT’s rapid text processing capabilities in less studied fields to produce impactful results, which underscores the importance of validating the methods before applying them to inform policy making in areas like K-12 public education.

Since the release of OpenAI’s latest model, GPT-4, multiple studies have observed improvements in performance and functionality over the previous models (e.g., GPT-3, GPT-3.5, GPT-3.5-turbo) \cite{lyu2023translating, huang2023benchmarking, sprenkamp2023large}. Moreover, the performance of GPT models is intricately linked to the design and phrasing of the prompts they are given \cite{cheng2023gpt, wang2023chatgpt, zhao2021calibrate}. Systematically designed prompts that specify instructions and tasks for GPT can lead to more accurate and focused responses, reducing randomness, oversimplification, and overlooked information, particularly in expertise-rich content. Carefully designed prompts may even mitigate some of the model’s biases \cite{gao2023collabcoder}.

Studies on how to enhance information processing using LLMs are proliferating with unprecedented speed. This paper contributes to the current literature by expanding the investigation to a rarely studied yet high-stakes field. Methodologically, this study demonstrates a comprehensive procedure for applying GPT-4 to facilitate qualitative textual analysis, aiming to minimize latent bias and attempt to bridge inherent field-specific knowledge gaps of the model through a grounded development of the codebook by established conceptual frameworks and domain expertise in prompt designs.

\subsection{Traditional NLP}
\subsubsection{LDA}
In the realm of traditional NLP, Latent Dirichlet Allocation (LDA) has emerged as a popular tool for the discovery stage of text analysis by extracting latent themes from texts \cite{leeson2019natural, jelodar2020deep}. Compared to LLMs, LDA maintains advantages in terms of transparency and interpretability, in addition to having well-documented operating procedures. The literature has documented a variety of strategies for validating unsupervised textual analysis results. These strategies often involve: (a) comparing the results with human expert coding of the same data, (b) juxtaposing the results with alternative data sources concerning the same phenomena, and (c) predicting criterion measures.

For example, Grimmer [2013]\cite{grimmer2013appropriators} applied an unsupervised model to analyze senators’ self-presentation to their constituents and then developed a codebook. A research assistant classified a portion of the documents according to this codebook, revealing a high correlation of 0.96 between human and computer coding \cite{grimmer2013text}. Similarly, Sun et al. [2019]\cite{sun2019using} utilized structural topic modeling to analyze textual data from over two million reform tasks that K-12 public schools in Washington State had designed and implemented. The results corresponded well with school leaders’ perceptions of reform priorities gathered through interviews and were significantly associated with reductions in student chronic absenteeism and improvements in student achievement.

Baumer et al. [2017]\cite{baumer2017comparing} used grounded theory---an interpretive qualitative method widely used in social science \cite{glaser1968discovery}---and topic modeling to analyze the same survey data. The results show that the two analyses produce some similar and some complementary insights about the phenomena of interest, in their study, the non-use of social media. This comparison underscores the potential for future research to investigate mixed computational-interpretive methods that synergize human coding and computer textual analysis to enhance knowledge discovery in social sciences.

In our study, we initially employed human qualitative coding combined with LDA topic discovery. The outcomes from the LDA were then used to inform the development of the final version of the codebook. This process included examining highly representative documents within topics and extracting high-frequency keywords. Additionally, we assessed the overlapping of LDA’s topic labels between the results of human coding and conducted a comparative performance analysis for LDA and GPT-4.

\subsubsection{Lexical-based Methods}

Lexical models represent one of the major approaches for sentiment analysis, with NLTK VADER being a popular lexical-based algorithm. At the time of its release, it met the benchmarks for state-of-the-art models \cite{hutto2014vader} and was noted for performing well with short texts, such as social media posts. It adeptly handles negations, such as “not good,” and adverb modifiers. Although it may not compare to the more recently developed state-of-the-art models, a recent study that applied seven sentiment analysis tools—Stanford, SVC, TextBlob, Henry, Loughran-McDonald, Logistic Regression, and VADER—to process social media posts and news articles found that VADER was still capable of outperforming the others \cite{das2021comparative}. Nonetheless, like other rule-based classifiers, VADER exhibits a weaker ability to recognize underlying tones and sarcasm in the absence of non-textual content such as emojis or social media tags \cite{gosavi2022transformer, nguyen2021sarcasm}.

\section{The Conceptual Framework: Resource Equity Policy}
\begin{figure}[h]
  \centering
  \includegraphics[width=0.7\linewidth]{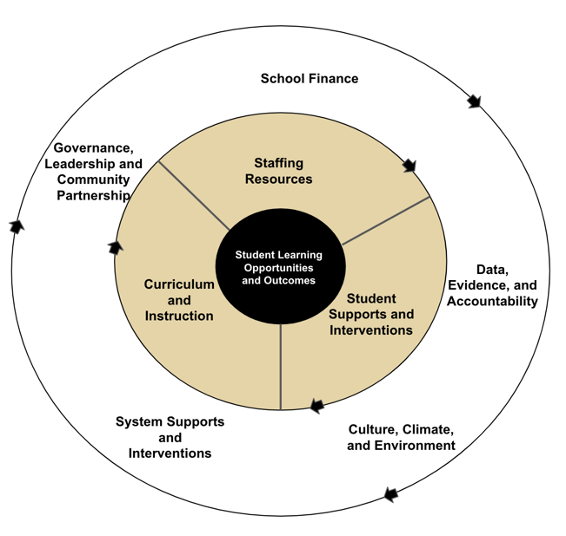}
  \caption{Resources Equity Framework Situated in a Data-Informed Iterative Improvement Cycle}
  \label{fig:framework}
  \Description{Resource Equity conceptual framework}
\end{figure}

Drawing on prior research in educational equity policy \cite{dimensions2023}, we conceptualize a Resource Equity framework that includes six essential components of educational policies that influence equitable student learning experiences and outcomes in schools (Figure~\ref{fig:framework}). The “inner circle” of policy strategies that are most proximate to students and have direct impacts on student learning includes: (1) the diversity and qualifications of \textit{school staff} (teachers and other adults) who have close interactions with students in schools; (2) the \textit{curriculum and instruction} that enable teachers and students to actively engage with rigorous and culturally relevant learning content; and (3) other types of \textit{student support and intervention} programs, such as mental health and social work services, multi-tiered support systems, summer school, and tutoring, which directly support students outside and around the classroom. 

The “outer circle” of support includes (4) \textit{school finance} that allocates resources to schools to support the offering of educational services, (5) \textit{school governance, leadership and community partnership} that determine school decision-making structure and power dynamics among stakeholders, (6) the \textit{data and evidence} that either enable or constrain the design, implementation, and evaluation of all the previous five components, as well as evidence-based accountability for effective and equitable use of educational resources, and (7) \textit{system supports and interventions} that include system-level reforms to better support students with academic and social-emotional needs, such as whole school improvement efforts. In addition, we acknowledge the (8) \textit{culture, climate, and local contexts} that constitute the environment for student and family experiences in- and outside of the school building. 

Policy and practices pertaining to each component at each level of the school system and across the hierarchy of schooling systems are embedded within a continuous cycle of improvement. In this iterative improvement cycle, it is critical to strategically incorporate stakeholder voices from diverse racial backgrounds, professional experiences, and geographic locations in the state.

\section{Data and Sample}
\subsection{Data Collection: Interview Process}
The data for this study comprise 24 interviews with diverse educational policy stakeholders. Our purposeful sampling strategy was designed to maximize representation based on the following criteria \cite{maxwell2004causal, patton1990qualitative}: (a) the level within public school systems, including classrooms, schools, districts, and the state; (b) geographic locations within the state; (c) roles of interviewees, spanning system actors at various levels of educational systems and three branches of the government at the state level, as well as non-system actors such as community organization leaders, advocates, lobbyists, teacher union representatives, and philanthropic organizational leaders; and (d) characteristics of students and local communities in terms of race/ethnicity, socioeconomic status, language, and homeless populations. 

The interviewees were categorized into three primary professional groups: policymakers and administrators (state legislators, state-level policymakers, and school district administrators), educators (teachers and those in coaching or mentoring roles), and non-profit sector participants along with advocates (including teacher union representatives, policy advocates, and community leaders). Interviews were conducted virtually via Zoom, each lasting between 45 to 60 minutes.

We intentionally designed the semi-structured interview questions to be broad, facilitating the emergence of a wide range of topics or deeper insights \cite{bhattacharya2017fundamentals} (see Online Appendix A.1 for the interview protocol). Interviewees were requested to provide examples of current state and local policies that they believed most significantly enhance or limit racial and economic equity in Washington state’s K-12 public education system. We further explored their reasoning and the mechanisms they proposed. Additionally, we solicited their perspectives on access to reliable data and evidence to support policy development and implementation, as well as their suggestions for iterative policy improvement at the state and local levels.

\subsection{Preprocessing Interview Data}
Audio recordings were transcribed into text. Interview data, unlike formal written language from documents or social media posts, often features loosely structured sentences, with interviewees frequently pausing mid-sentence, using filler words like “um,” “so,” and “you know,” or switching topics abruptly. Our research assistants meticulously listened to all 24 audio recordings, making necessary corrections to the transcribed texts and sentence structures. After the initial cleaning process, we compiled the data into a tidy text format, resulting in approximately 1,400 entries (i.e., paragraphs). Each paragraph encapsulates one complete thought, which may span one or several sentences. Although the text corpus is not “big” in the computer-assisted textual analysis literature, it provides sufficient data for computational methods to yield meaningful results without making the iterative human coding process intractable. The dataset also includes variables such as interviewees’ research identification, demographics, job roles, and job locations.

\section{Method: Human-Computer Interactive Learning}
\begin{figure}[h]
  \centering
  \includegraphics[width=\linewidth]{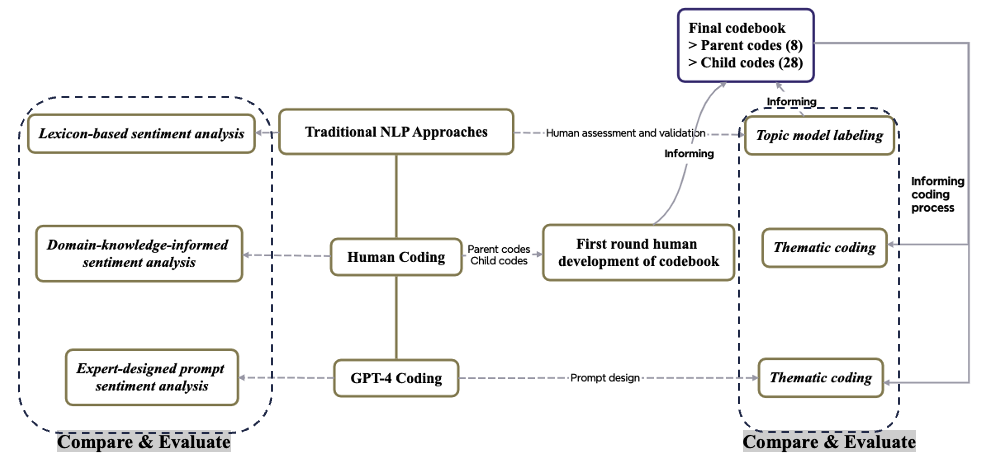}
  \caption{Workflow Diagram}
  \label{fig:workflow}
  \Description{This figure shows the flow of our proposed framework utilizing human expertise, GPT-4, and traditional NLP approaches for thematic and sentiment analysis. }
\end{figure}

Our methods incorporate multi-stage interactions between human and computer to conduct both qualitative and quantitative analysis to examine these four research questions. Figure \ref{fig:workflow} summarizes the key aspects of the overall workflow, showing the flow of our proposed framework utilizing human expertise, GPT-4, and traditional NLP approaches for thematic and sentiment analysis. 

\subsection{Codebook Development}
\subsubsection{First Round of Human Qualitative Coding for Initial Codebook Development}
Three qualitative coders, all with education policy research knowledge and extensive experience working in K-12 schools, participated in the first round of coding to familiarize themselves with the interview data and to generate an initial codebook. One coder employed a grounded theory approach, iteratively developing a codebook informed by the Resource Equity framework as stated in Conceptual Framework section and detailed reading of 10 interview samples. This initial codebook was then used by three expert coders to code the interview data and conduct qualitative analysis for another study within the larger project, which confirmed the content validity of our initial codebook and conceptual framework.

\subsubsection{LDA}
Concurrent with the first round of human qualitative coding, we employed LDA to uncover themes and patterns through latent semantic analysis of word and phrase probability distributions. The analytical process, including the selection of the optimal number of topics, topic labeling, and cross-validation, adhered to established practice guidelines (detailed information on topic modeling analysis can be found in Appendix B). The topics were labeled in accordance with the Resource Equity framework. To manually validate the themes identified by the topic modeling, we developed rubrics as detailed in Appendix A.2. This validation involved scrutinizing 20 documents with the highest top proportions and the 10 most frequently occurring words to interpret the topics’ meaning and coherence. Out of 30 topics identified, 25 were deemed both theoretically coherent and practically relevant to the policy interests within the context of Washington State.

\subsubsection{Code Refinement and Final Codebook Creation}
Subsequently, we integrated the themes from both the human-developed codebook and the LDA findings to construct the final codebook. The initial codebook was refined with the help of LDA, which provided more structured code labels and identified high-frequency keywords for the child codes. The finalized codebook contained eight broad parent codes that reflect the major themes from the stakeholder interviews and are aligned with our conceptual framework, thereby guarding the content validity of the text analysis by ensuring its results encompass the concepts pertinent to the policy issue of interest. These parent codes include (1) culture, climate, and environment; (2) curriculum and instruction; (3) data, evidence, and accountability; (4) governance, leadership, and community partnership; (5) school finance; (6) staffing resources; (7) student supports and interventions; and (8) system supports and interventions. Within these parent codes, we created 28 child codes to represent specific topics of discussion. The codebook, detailed in Appendix A.3, lists parent and child code labels, descriptions, and keywords.

We opted for topic modeling over LLMs at this juncture for several reasons. Firstly, to avoid the potential biases inherent in algorithmic coding during the inductive coding phase, as we sought to establish baseline results for automatic analysis using the codebook. Secondly, the topic classification and labeling with LDA preceded the final refinements to the codebook; hence, subsequent changes to the codebook did not affect the LDA outcomes. In contrast, if we incorporated GPT-informed codes into the codebook, those codes would be used for GPT-4 labeling and might have unduly favored it in thematic analysis. Furthermore, the straightforward interpretability of LDA facilitated dimensionality reduction of unstructured text data, allowing for model improvement through parameter adjustments. Although LDA’s simplicity might overlook subtleties that LLMs could capture, our human coders compensated for this with their nuanced understanding and domain-specific reasoning. Conversely, LLMs could introduce unnecessary complexity at this stage of pattern discovery, such as computational intensity and potential over-interpretation of text, making LDA a more appropriate choice for supporting discovery without adding undue complexity.	

\subsection{Thematic Annotation}
\subsubsection{Human Thematic Coding as Ground Truth}
To ensure the validity and consistency of the human annotation as the benchmark for evaluating machine-generated results, we conducted two rounds of extensive coding and involved coders without a machine learning background. Once a codebook was developed, two doctoral research assistants were trained to annotate the entire interview dataset. These assistants, who were not involved in the codebook development and who had substantial training in educational policy in Washington state, used the codebook to identify up to three most salient themes for each paragraph from child codes. If no appropriate child code was applicable, they were to select the most fitting parent codes. 

During the training phase, both coders independently annotated a shared set of 50 paragraphs. They achieved consensus on over 75\% of the coding, with any inconsistencies resolved through discussions aimed at harmonizing code application. Meetings were instituted to facilitate continuous dialogue, address uncertainties, and promote uniform coding practices. Having attained a stable inter-rater reliability, they coded the entire corpus and reported no cases of paragraphs containing more than three themes. The second round of human annotation served as verification. Liu and Sun, who were familiar with the interview data and the codebook, reviewed and revised the coders’ annotations to increase alignment with the codebook. This process yielded 0-3 codes for each unit of analysis, with 12\% at the parent code level and 88\% at the child code level. The codes identified by human coders were later used to compare with the topics assigned by the machine for each document.

We used human annotations to address substantial research questions. For SRQ 1, we analyzed the frequency of topic appearance during the interviews and further summarized the themes by stakeholders’ job roles. We hypothesized that teachers and teacher mentors would discuss topics such as staffing resources, recruitment, retention, and professional development more frequently due to their direct experiences and relevant knowledge. District and state administrators were expected to focus on issues like school finance and resource allocation, whereas non-profit organizations would likely emphasize the involvement of parents and communities in policymaking. If the analysis confirmed our hypotheses, it would add credibility to the human versus machine annotation comparison.

\subsubsection{GPT-4 Thematic Annotation}
To achieve optimal results from GPT-4, we developed multiple prompts, adjusting for the number of steps, examples, and instructions. We conducted several rounds of prompt testing using different parameters, manually reviewing the outcomes and the logic of the responses.

\paragraph{Prompt designs} 
Building on recent studies in prompt engineering, we designed and tested both zero-shot and chain-of-thought (CoT) prompts to reflect our double-layered codebook structure \cite{liu2021p, wei2022chain, kojima2022large}.

\begin{itemize}
    \item Zero-shot: use the codebook to label three child and/or parent codes for each document; if no child code works, only label the parent codes.
    \item CoT: Step 1: label three parent codes with reasoning; Step 2: label child code within each parent code. If none of the child code applies, keep the parent code.
\end{itemize}

The instructions for GPT-4 mirrored those given to human coders, with the additions of study context and GPT’s role to bridge the information gap between GPT-4 and the coders. For all prompt variations, GPT-4 was specifically instructed to consider the context of the Washington State K-12 public school system. After extensive testing, we found that CoT prompts yielded a higher alignment with human annotations (54\% for zero-shot vs. 77\% for CoT), consistent with findings that CoT prompts enhance GPT’s performance \cite{wei2022chain, kojima2022large}.

\paragraph{GPT Analysis Settings} GPT-4 was set to a temperature of 0.5. GPT temperature is a setting that controls the randomness of responses, ranging from 0 to 1 with a higher temperature allowing for more varied and creative outputs, while a lower temperature results in more predictable and conservative answers. Previous studies employing GPT for semantic annotation often selected a temperature of 0 to ensure reproducibility and low randomness (e.g., \cite{dai2023llm, xiao2023supporting, chew2023llm}), albeit at the cost of limiting GPT’s exploratory capabilities. Since no definitive guideline on temperature setting exists, we experimented to find the optimal setting for our data. A temperature of 0.5 struck a balance between coherence and diversity in the GPT generated results, ensuring responses are reasonably predictable and on-topic, while still allowing for a moderate level of variation \cite{openai2023cheatsheet}. To address concerns about inconsistent results at this higher temperature, we employed three strategies: providing GPT with detailed task descriptions, verifying reproducibility on a subset of 50 randomly sampled paragraphs, and evaluating the randomness by comparing agreement rates with those calculated from shuffled labels. Our tests showed over 98\% of CoT results were reproducible and the shuffled agreement rates dropped substantially from 77.89\% to 17.89\%, indicating that the GPT labels were paragraph-specific rather than randomly assigned. 

We refined the prompt instructions and tested them on randomly sampled data. Upon reviewing the codes and reasoning from the prompt variants, we selected the CoT prompt format as detailed below:

\noindent\fbox{
    \parbox{0.95\textwidth}{
prompt\_base = f"""
Task: As a policy researcher, you’ve been provided with a paragraph extracted from an interview with an education policy stakeholder. Utilize the provided Codebook (in CSV format) to code the paragraph. The Codebook comprises four columns: ‘Parent’, ‘Child’, ‘Child\_description’, and ‘Key words’.

Steps:

1. Identify Salient Themes:
\begin{itemize}
    \item Understand the paragraph’s content within the context of the Washington State K-12 public school system.
    \item Refer to the ‘Parent’ column in the Codebook for broader thematic categories.
    \item Pinpoint up to three salient themes from these ‘Parent’ categories.
    \item These themes should highlight the most significant ideas in the paragraph.
    \item Label the paragraph with the chosen ‘Parent’ themes.
\end{itemize}

2. Dive into Child Themes:

\begin{itemize}
    \item The ‘Child’ column in the Codebook lists detailed thematic subcategories, which fall under the broader ‘Parent’ categories.
    \item The ‘Child\_description’ elaborates on the ‘Child’ categories, and the ‘Key words’ column lists pertinent terms for each ‘Child’ category.
\end{itemize}

3. Associate with Child Categories:

\begin{itemize}
    \item Revisit the paragraph, keeping the Washington State K-12 public school system context in mind.
    \item For each previously identified ‘Parent’ theme, pinpoint the appropriate ‘Child’ subcategories from the Codebook. The ‘Child\_description’ and ‘Key words’ columns can aid your decision.
    \item Ensure the ‘Child’ categories align with the paragraph’s content. If there’s no fit or you’re uncertain, label it as ‘None’.
    \item From your identified ‘Parent’ and ‘Child’ pairs, pick the top three pairs that encapsulate the paragraph’s central ideas.
    \item Label the paragraph with these three ‘Parent’ and corresponding ‘Child’ pairs.
\end{itemize}

Codebook:
\{codebook\}

Paragraph for Analysis:
[[[TEXTGOHERE]]]

Response Format:
Frame your answer as a JSON object containing the keys: ‘Parent 1’, ‘Child 1’, ‘Parent 2’, ‘Child 2’, ‘Parent 3’, ‘Child 3’, and ‘Reasoning’.
"""
}}

In this prompt, we divided the annotation task into three parts to facilitate GPT-4’s step-by-step thinking. We provided the codebook to GPT-4 (codebook text replaced \{codebook\}) and fed it one paragraph at a time (input paragraph replaced “TEXTGOHERE”). First, in the predefined context as the specified role, GPT-4 identified three most salient broad themes for each paragraph only using information in the “Parent” code column of the codebook ($p_1$, $p_2$, $p_3$). Then, GPT-4 learned columns in the codebook that related to child codes. Subsequently, GPT-4 utilized the learned information from “Child” code columns to enumerate suitable child codes under selected parent codes ($p_i c_j$, $i \in \{1,2,3\}$, $j \in \{i,...,n\}$). In this step, we requested GPT-4 to use exemplary high-frequency keywords and found that this request improved the GPT-4 performance. After attaining a collection of parent-child pairs, we asked GPT-4 to select three pairs of the most salient themes for the paragraph based on the child code description. If GPT-4 viewed that a given paragraph did not necessarily contain three Child Codess, it could return “None” to fill in the spot for the absent label. For some paragraphs, GPT-4 was unable to identify proper child codes, so it followed the instruction to choose parent codes instead, which mirrored human coders’ thought process and workflow. The output from this prompt gave us up to three parent codes and their corresponding child codes, along with GPT-4 reasoning for selecting these codes. Here is an example of GPT-4 output:

\noindent\fbox{
    \parbox{0.95\textwidth}{
{"Theme 1": "Accountability system",

"Theme 2": "Instructional programs",

"Theme 3": "Teacher union, salary, workforce",

"Reasoning": "The paragraph discusses the need for accountability in ensuring equal delivery of instruction across different regions (north and south side), which corresponds to the 'Accountability system' theme. It also mentions the delivery of instruction, which fits the 'Instructional programs' theme. Finally, the speaker refers to potential solutions that the union might not accept, which relates to the 'Teacher union, salary, workforce' theme."}
}}

\subsubsection{LDA}
The LDA results, including topic proportions and topic labels, were obtained prior to the finalization of the codebook. To harmonize with the structure of human annotation, we assigned three codes to each paragraph based on the highest topic proportions. These were classified as child code labels. As the machine did not directly assign parent codes, and because we could only identify labels for 25 out of the optimal 30-topic model, there were three child codes\footnote{Child codes---including data capacity, accountability system, and instructional programs were not identifiable during LDA topic labeling.} from codebook did not appear in the LDA results (as detailed in Appendix B).

\subsubsection{Parent and Child Code Level Analysis}
The codebook’s hierarchical structure led to an evaluation of machine performance at both broad (parent code) and specific (child code) theme levels. This involved standardizing annotation outputs, mapping child codes to parent codes, and removing duplicates, resulting in 0-3 codes at each level per paragraph. The methodological research questions were addressed by analyzing outcomes using both child and parent codes, where “original” labels refer to unmodified human or machine-assigned codes, and “parent codes” refer to all parent-level codes including those mapped from their child codes.

\subsection{Sentiment Annotation}
\subsubsection{Human Sentiment Annotation}
Liu and Sun conducted sentiment annotations. To ensure objectivity and content validity, they closely followed established guidance. They labeled the sentiment expressed in each paragraph as “Positive,” “Neutral,” or “Negative.” “Positive” labels were assigned when interviewees expressed satisfaction with a policy or practice, demonstrated improvement from past practices, or identified policies or practices that have enhanced or have the potential to enhance educational equity. Conversely, “Negative” labels were used when interviewees expressed dissatisfaction, identified issues or challenges, or demanded improvements. When a document/text merely describes the fact without expressing either “Positive” or “Negative,” we code it “Neutral.” After establishing common understanding of the definitions, the two researchers coded the interviews independently and then cross-verified each other’s coding. To maintain consistency throughout the coding process, the two coders discussed any documents/texts they were unsure of or had differential coding or understanding to reach consensus. Specific child codes in the codebook, such as “Progressive funding” and “Tests and inconsistent standards for college readiness and student success,” were recognized as inherently carrying sentiment in the context of studying policies advancing educational equity. Hence, we hypothesized a higher occurrence of negative labels associated with “Tests and inconsistent standards for college readiness and students’ success” and a lower frequency of negative labels for “Progressive funding” in the sentiment analysis results. This difference was anticipated to serve as a verification of the validity check for the sentiment classification.

\subsubsection{GPT-4 Sentiment Annotation}
The prompt for GPT-4 was carefully designed to reflect the task performed by the human annotators, following the same guidelines:

\noindent\fbox{
    \parbox{0.95\textwidth}{
prompt\_base = f"""
Act as a policy researcher, you will classify the sentiment in the interviews of educational policy stakeholders as: “Positive”, “Negative”, or “Neutral”. Here is a statement from a policy stakeholder: 

[TextGoHere]

To warrant “Positive” sentiment, the statement has to: (1) include the interviewee’s satisfaction about an educational policy (policies) and program(s), or (2) express an enhancement or potential to enhance the quality or equity of student learning or school system, or (3) identify an improvement from past practice. To warrant “Negative”, the statement describes the interviewees’ dissatisfactions, or identifies problems/issues/challenges, or suggests areas needed for further improvement. When the interviewee just states the fact without expressing either positive or negative sentiment, you can classify as “neutral”. When multiple sentiments are observed in one statement, identify the most prevailing sentiment. Explain your reasoning for your analysis."""
}}

The output from GPT-4 included the sentiment label and a brief reasoning for assigning that label. Here is an example output:

\noindent\fbox{
    \parbox{0.95\textwidth}{
{"Sentiment": "Negative",

"Reasoning": "The interviewee expresses dissatisfaction with the tendency for teachers to delay teaching linear equations until February, resulting in a three-month gap in the curriculum. They believe that kids could learn them anytime with proper scaffolding, implying that the current practice is not effective or efficient. This statement highlights a problem or challenge in the educational system, warranting a negative sentiment classification."}
}}

\subsubsection{Lexical-based Sentiment Analysis}
We utilized the nltk.sentiment.vader package in Python for lexical-based sentiment analysis. This tool is one of the commonly used, best-performing lexical-based tools for sentiment analysis \cite{abdulaziz2021topic, belal2023leveraging, das2021comparative}. The interview text was input into the pre-computed algorithm, which then associated words with sentiment scores and compounded the overall sentiment scores for paragraphs, taking into account negations and intensifiers. These compound sentiment scores, ranging from -1 (most negative) to +1 (most positive), were then classified into “Positive,” “Negative,” and “Neutral” sentiment labels based on default cutoffs\footnote{Positive: If the compound score is $\geq$ 0.05. Negative: If the compound score is $\leq$ -0.05. Neutral: If the compound score is between -0.05 and 0.05.}. These cutoffs were designed to be generally effective across various texts, as determined through extensive testing and validation by the developers of the tool.

\subsubsection{Comparing GPT and traditional NLP to Human Coding}
We navigated the proposed research questions using various analyses. For the substance research questions, we applied semantic analysis to explore the underlying themes and sentiments in the interview data and discovered unique patterns that aligned with the interviewees’ job roles and responsibilities. The patterns were consistent with our hypotheses, which were derived from education policy literature and authors’ understanding of WA policy contexts. These findings cast credibility to the human annotation results as the baseline for assessing the validity of computer-assisted analysis results. 

To examine the validity of GPT-4’s thematic and sentiment analysis in the WA state K-12 public education arena, we conducted systematic evaluations across different dimensions. We assessed thematic agreement using overlap metrics and compared the machine’s coding with human analysis results using confusion metrics. We then proceeded to evaluate topic-level alignment between machine and human codings using bootstrapped reliability measures for label similarities and cosine similarities with tf-idf for text similarities. These measures allowed us to gain a comprehensive view of the construct and criterion validity of using GPT-4 for text analysis in our domain context.

\section{Results}
In this section, we summarize key findings by each research question. 

\subsection{SRQ 1. What are the key themes that WA stakeholders voiced about the K-12 public school system?}

Human-annotated code frequencies (original labels) revealed the primary areas of the WA K-12 public school system as highlighted by stakeholders during their interviews. Table~\ref{tab:human_freq} shows stakeholders’ extensive focus on data related topics, such as data collection and access, and use of data to help practitioners improve practices and inform policy making. This also includes issues of data sharing, reporting, transparency, and quality (“Data access, analysis, reporting, use, quality and transparency”). Beyond data-related concerns, inclusive governance in the K-12 public education system is one of the central topics, with four of the top five frequent themes related to it. Stakeholders emphasized the importance of meaningful engagement, such as building relationships and centering voices of the marginalized youth and families (“Coalition and relationship”) and collaborating with their communities (“Community”). In addition to bringing marginalized communities actively into the conversation, they also advocated for increasing representation of these communities in leadership roles (“Leadership in diversity”) to enhance diversity, inclusivity, and anti-racism in the system. Given that the interviews were situated in the context of the K-12 public school system, it was not surprising that “Governance, leadership, and community partnership” emerged as a salient topic. 

\begin{table}
\begin{threeparttable}
\caption{Human Label Frequency}
\label{tab:human_freq}
\begin{tabular}{p{0.3\textwidth} p{0.1\textwidth} p{0.3\textwidth} p{0.1\textwidth}}
\toprule
Parent Codes (aspects in parent codes that are not covered by child codes) & Parent Code Frequency (\%) & Child Codes & Child Code Frequency (\%) \\
\midrule
Data, evidence, and accountability & 0.32 & Data access, analysis, reporting, use, quality and transparency & 9.32 \\
Governance, leadership, and community partnership & 4.64 & Coalition and relationship & 6.77 \\
Governance, leadership, and community partnership & 4.64 & Community & 5.89 \\
Governance, leadership, and community partnership & 4.64 & Leadership in diversity & 5.38 \\
Culture, climate and environment & 1.16 & Anti-racism & 5.06 \\
Staffing resources & 0.19 & Diversify teacher workforce (teacher labor market) & 4.78 \\
Student supports and interventions & 1.11 & Learning opportunities and programs & 4.31 \\
Staffing resources & 0.19 & Mentoring, coaching, and teacher learning & 4.08 \\
Data, evidence, and accountability & 0.32 & Goals, outcomes, and measures: Tests, standards, graduation requirements & 3.57 \\
System supports and interventions* & & School system support and improvement & 3.53 \\
Curriculum and instruction* & & Curriculum development and instructional delivery & 3.34 \\
School finance & 1.35 & Funding formula & 3.06 \\
Governance, leadership, and community partnership & 4.64 & Legislation process & 2.78 \\
Staffing resources & 0.19 & Teacher union, salary, workforce & 2.74 \\
Governance, leadership, and community partnership & 4.64 & Local control and district policies and politics & 2.64 \\
Data, evidence, and accountability & 0.32 & Accountability system & 2.55 \\
Student supports and interventions & 1.11 & Differentiated student strategies & 2.5 \\
Student supports and interventions & 1.11 & Multilingual programs & 2.27 \\
Governance, leadership, and community partnership & 4.64 & Government relationships & 2.18 \\
School finance & 1.35 & Targeted funds & 2.18 \\
Student supports and interventions & 1.11 & Students' SEL and health & 1.99 \\
Culture, climate and environment  & 1.16 & Trauma at home & 1.76 \\
Governance, leadership, and community partnership & 4.64 & School board & 1.67 \\
School finance & 1.35 & Progressive funding & 1.58 \\
System supports and interventions* & & Judicial systems & 1.58 \\
Curriculum and instruction* & & Instructional programs & 1.44 \\
Data, evidence, and accountability & 0.32 &Tests and inconsistent standards for college readiness and students' success & 1.16 \\
Data, evidence, and accountability & 0.32 & Data capacity & 1.11 \\
\bottomrule
\end{tabular}
\begin{tablenotes}
\item[*] All parent code frequencies were measured for original labels rather than converted parent codes from child codes. Parent codes indicated by * were not directly assigned to a paragraph by human coders.  
\end{tablenotes}
\end{threeparttable}
\end{table}

\begin{figure}[h]
  \centering
  \includegraphics[width=\linewidth]{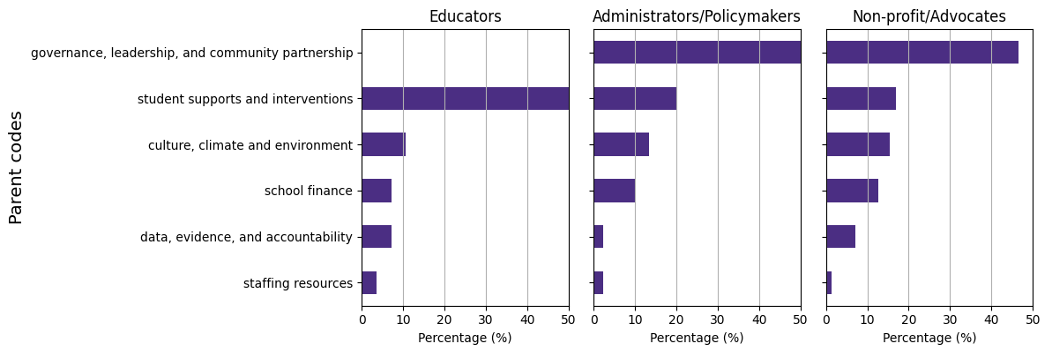}
  \includegraphics[width=\linewidth]{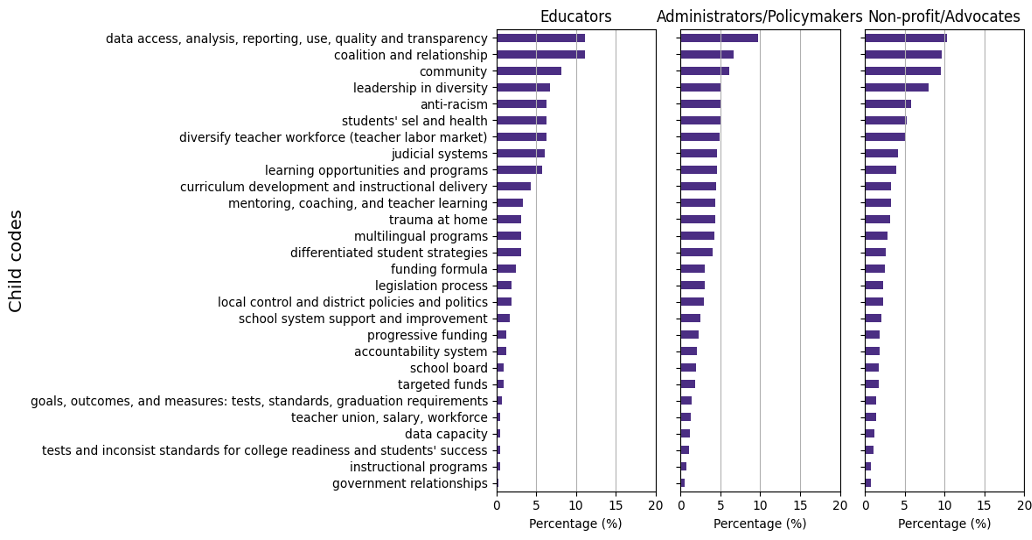}
  \caption{Human Labeled Theme Frequency by Stakeholders’ Job Roles}
  \label{fig:hc-freq}
  \Description{distribution of topics appear in stakeholders' interviews}
\end{figure}

When we disaggregated our analysis by stakeholders’ job roles\footnote{Our data includes three broad job role categories: (1) Administrators/Policymakers: state legislators, other state-level policymakers, school district administrators; (2) Non-Profit and Advocates: teacher union representatives, policy advocates, and community leaders; (3) Educators: teachers, teacher coaches or mentors.}, we developed a more nuanced understanding of the themes. As illustrated in Figure~\ref{fig:hc-freq}, the human annotation results largely indicated the same patterns of thematic distribution among the three types of stakeholders’ job roles: administrators and policymakers, educators, and non-profit advocates. On the other hand, these three types of stakeholders voiced unique concerns related to their daily work, expertise, and lived experience. For example, educators, including teachers and their mentors, concentrated more on “Staff resources,” such as teacher education for diversifying the teacher workforce (“Diversifying the teacher workforce and teacher labor market”), and “Student supports and interventions,” including learning opportunities in schools and access to curriculum programs (“Learning opportunities and programs”). Administrators and policymakers discussed “School finance” related themes, including “Targeted funds,” “Progressive funding” practices, “Funding formula” revisions, along with their work around bills and legislative process for educational policies (“Legislation process”). Nonprofit advocates highlighted governance and community relationships.

\subsection{SRQ 2. Which themes did stakeholders recognize as advancing educational equity (positive)? Conversely, which areas were mentioned as needing improvement or hinder (negative) educational equity?}

\begin{figure}[h]
  \centering
  \includegraphics[width=\linewidth]{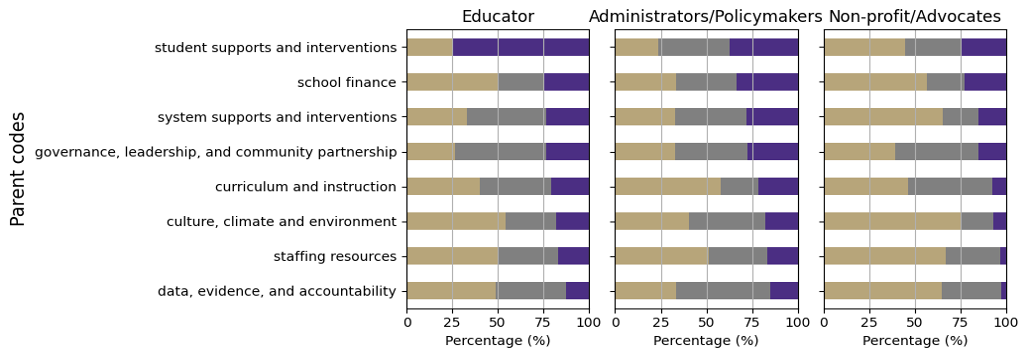}
  \includegraphics[width=\linewidth]{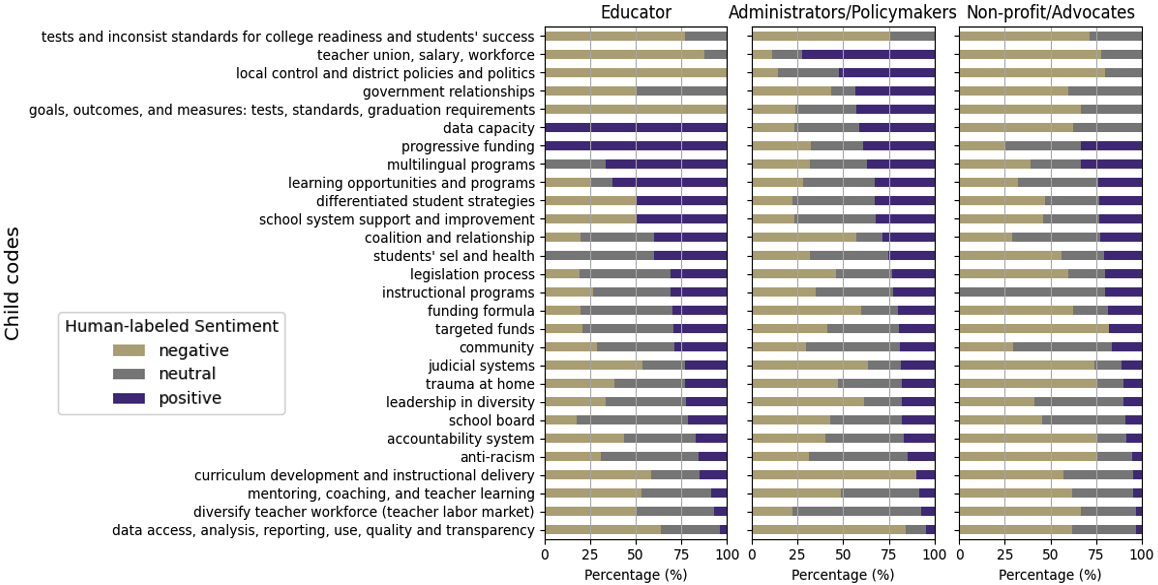}
  \caption{Human Labeled Sentiment by Stakeholders’ Job Roles}
  \label{fig:hc-sentiment}
  \Description{distribution of sentiment appear in stakeholders' interviews}
\end{figure}

The study found that very few areas in the WA state K-12 public education system received a majority of positive sentiment, which might highlight a need for improvement across many aspects of the system. Figure~\ref{fig:hc-sentiment} illustrated the proportion of positive, negative, or neutral sentiments that WA state K-12 public education stakeholders expressed within each topic area, as delineated by child or parent codes. 

A larger proportion of positive sentiment was expressed about progressive funding. Additionally, stakeholders commonly acknowledged the efforts and preliminary positive results in student support and interventions, particularly appreciating the reform of multilingual programs. The reform switched previously adopted late-exit programs, which focused on transitioning ELL students from home languages to English, to the multilingualism programs that committed to honor students’ cultural and linguistic heritage. As a local administrator illustrated:

\begin{quote}
    I think the thing that we have been able to do in that shifting, in that transition, is really clarify our commitment to bilingualism. And especially for our families who are second language, around that is your heritage language, that is the language of your ancestors. That is what connects us to who we are and the generations who came before us, and how important that is.
\end{quote}

Such transitions demonstrated WA’s K-12 system’s commitment to equitable learning opportunities, differentiated support for students with diverse needs, and culturally responsive program design. Nonetheless, stakeholders also pinpointed persistent issues such as inadequate student social-emotional learning and health support, attributed to lack of funding and insufficient staff resources—issues that became pronounced during the remote learning mode of the COVID-19 pandemic. A representative from a statewide nonprofit organization indicated the dire state of student mental health resources: “it is important to note the lack of necessary mental health resources for students around the state [...] and COVID-19 has raised the impact that the mental health crisis have been having on our students due to the lack of resources available.”

In addition to staff resources, the state’s data, evidence, and accountability system was critiqued for several commonly recognized concerns. Many stakeholders spoke positively about the volume of data collected and available within the state data system. However, they highlighted a distinction between data collection and data accessibility, critiquing the system’s complexity for creating barriers to those lacking data capacity and network connections. Furthermore, inconsistent standards and outcome measures, coupled with an absent accountability system, led to confusion and gaps in feedback loops across various aspects of the system. For instance, this lack of clarity adversely affected educators’ ability to deliver an equitable curriculum, as one stakeholder noted, “there should be some accountability that a kid on the north side is going to get the same delivery of instruction as on that south side. And I think that is still a gray area for schools.” Additionally, the absence of unified standards and accountability posed challenges in monitoring funding allocations for state administrators, with one commenting, “when you are looking at funding, we have put in almost \$10 billion in the last 10 years, but there is really no accountability system to that. We still have 23 different accounting systems that feed into OSPI, and then they have to figure it out.”

Stakeholders’ attention to various areas reflected their roles and responsibilities, leading to divergent perceptions of the same issues. Compared to their counterparts in other roles, stakeholders from non-profit organizations and advocacy groups expressed general dissatisfaction with the current practices and policies and called for improvements in numerous areas. This dissatisfaction often stemmed from the system’s inequitable treatment of marginalized students, families, and communities. A non-profit representative shared her observation of racial stratification within the system, recounting an instance where black parents moving into a white community faced discouraging and demeaning challenges. She recalled a parents-teacher conference where the question posed to the parents was a telling one: “What makes you qualify?” She stressed that “we cannot just ask black and brown people to enter spaces like that without some pretty intense incentive.”

These barriers impacted not only community and family access to the system but also influenced decisions related to the recruitment and retention of teachers of color. This caught the attention of state administrators aiming to diversify the teacher workforce and of representatives of teachers of color. Institutional and systemic racism led to a situation where, as an administrator acknowledged, “there are very few teachers that stay for longer than five years. And a lot of them are exhausted. And a lot of the folks that we talk to, they leave because they do not want to be in the racial trauma of being in the school building, educating our kids and trying to decolonize the educational curriculum at the same time.”

Educators recognized the limitations of policies concerning teacher unions, salaries, and the workforce. They also articulated a need for policies and practices to be designed and implemented with the aim of supporting students’ academic and social-emotional learning. Furthermore, practitioners suggested that the goals of such policies should be evaluated during implementation, rather than simply requiring compliance under the umbrella of local district control or complex government relationships. As one educator put it, “I know that a couple of years ago the district came out with an equity policy. I think what is very interesting is how that actually plays out in leadership in classrooms. It felt like a very formal or not formal, but just like a checkbox. We did the thing, we wrote the thing we are going to abide by these policies that are very vague and not very specific.”

\subsection{MRQ 1. How accurate and valid are GPT-4 labels of key themes when comparing to human experts’ labels and traditional topic modeling results?}

\subsubsection{Evaluation for GPT-4 Thematic Analysis}
To evaluate the overlap between machine and human coding, we calculated hit rates, which measured the percentage of machine-labeled themes that corresponded with human-labeled themes. Table~\ref{tab:metrics_comparison} illustrates that, on average, approximately 78\% of GPT-4 annotated child codes matched those annotated by humans for each paragraph. Given that up to three child codes were assigned to each paragraph, these results indicate a significant overlap at child code level about detailed themes, with GPT-4 and human coders identifying at least two identical themes per paragraph. The LDA approach also demonstrated a high rate of overlap, aligning with human annotations for more than half of the child codes. Hit rates for both computer-assisted methods increased when comparing the broad themes (parent codes). In particular, GPT-4 parent code labels nearly fully cover the human experts’ coded parent codes. 

To address potential randomness in GPT-4 outputs when the temperature setting is higher than 0, we calculated shuffled hit rates by comparing machine annotations for a paragraph with human annotations from another randomly selected paragraph. This approach aimed to determine whether the computer-assisted methods might assign randomly generic themes. As shown in column 3 of Table~\ref{tab:metrics_comparison}, the decrease in shuffled hit rates suggest that both GPT-4 and LDA were capable of discerning paragraph content and assigning labels based on specific content. To test the robustness of our evaluation to different evaluation metrics, we also calculated Szymkiewicz-Simpson coefficients, S\o rensen-Dice coefficients, and Jaccard similarity indices. The results of these measures, presented in Appendix C Table~\ref{tab:c1robust}, are consistent with our findings in Table~\ref{tab:metrics_comparison} that GPT-4 surpassed LDA in accuracy for both child and parent level annotations.

\begin{table*}
\centering
\caption{Comparison of Agreement and Confusion Matrix Metrics}
\resizebox{\textwidth}{!}{
\begin{tabular}{lcccccccccc}
\toprule
& \multicolumn{5}{c}{Agreement Metrics (Child codes)} & \multicolumn{5}{c}{Confusion Matrix Metrics (Parent codes)} \\
\cmidrule(lr){2-6} \cmidrule(lr){7-11}
& \% Hit rates & \% Shuffled hit rate & Precision & Recall & F-Score & \% Hit rates & \% Shuffled hit rate & Precision & Recall & F1-Score \\
\midrule
GPT-4 vs. Human & 77.89 & 17.89 & 0.33 & 0.63 & 0.42 & 96.02 & 56.67 & 0.52 & 0.87 & 0.62 \\
LDA vs. Human & 60.65 & 13.66 & 0.23 & 0.38 & 0.27 & 76.13 & 47.80 & 0.43 & 0.64 & 0.49 \\
\bottomrule
\end{tabular}
} 
\label{tab:metrics_comparison}
\end{table*}

Furthermore, we assessed GPT-4’s annotation performance using measures derived from the confusion matrix. Notably, at both the child and parent code levels, GPT-4 and LDA exhibited high recall rates compared to precision. This implies that machine-annotated false negatives were uncommon, whereas false positives were more frequent. In other words, themes identified by human coders were likely to be recognized by the machines, but not all themes suggested by the machines were confirmed by the human annotators. The relatively low precision could result from the different output formats in that: human coders could choose between zero to three labels per paragraph flexibly; in contrast, the algorithms constrained the LDA to always assign three themes and GPT-4\footnote{Our prompt instructed GPT-4 to identify the three most salient themes, except in cases where there were not enough suitable themes available.} to identify three salient themes in most cases. Therefore, machine annotations naturally included more labels, leading to an increase in false positives. Nevertheless, the high recall rates confirmed claims from agreement metrics, illustrating that GPT-4 was adept at capturing human-identified themes and outperformed LDA in this respect. Additionally, the low precision might indicate that GPT-4 was able to uncover nuances in text analysis that had not been detected by human coders.

The agreement metrics and confusion matrix indicators present a comprehensive view of the overall performance of GPT-4 and LDA. Our interview data consisted of loosely structured but context-rich and information-dense texts, which featured multiple themes. Consequently, machines’ performance varied across different thematic contents. We applied bootstrapped Cohen’s $\kappa$, AUC, and cosine similarity with \textit{tf-id} to capture the differentiation on a code-wise level. To adapt these measures for binary classifications to our multi-label classification scenario, we transformed the current data using one-hot encoding \cite{dahouda2021deep}. Subsequently, we generated a binary dataset where each row represented a paragraph and each column corresponded to a code. A cell was assigned a value of 1 if the paragraph was associated with the code in that column, and 0 otherwise. Using 100 bootstrapped iterations, we obtained Cohen’s $\kappa$, AUC values, and corresponding 95\% confidence intervals for each code by comparing human and machine annotations.

Cohen’s $\kappa$, which accounts for the probability of chance agreement \cite{cohen1960coefficient}, was used to measure the inter-rater reliability between the computer-assisted models and human coding. Due to the nature of the interviews, where each theme constituted only a small part of the conversation, zeros were more frequent than ones, leading to an imbalanced dataset. In such cases, Cohen’s $\kappa$ is considered more robust than accuracy. As indicated in Table~\ref{tab:bootstrapped}, GPT-4 generally outperformed LDA on average, at individual parent code level, and for the majority of child codes (Appendix C Figure~\ref{fig:c2}-\ref{fig:c5}). The average Cohen’s $\kappa$ at the child level was comparable to deductive coding using an expert-informed codebook in linguistics \cite{xiao2023supporting} and was better than deductive coding using a GPT-informed codebook for an open-access dataset \cite{dai2023llm}. 

Our parent-level statistics showed even better performance. When comparing GPT-4 to human annotations for each code, Cohen’s $\kappa$ exceeded 0.4 for five parent codes and surpassed 0.7 for two parent themes “Staffing resources” and “School finance.” Although Cohen’s $\kappa$ for LDA and human annotations for child codes ranged between 0-0.5, the agreement on the theme of “Data access, analysis, reporting, use, quality, and transparency” exceeded 0.6, outperforming the agreement between GPT-4 and human for this theme. For the remaining child codes, GPT-4 demonstrated higher concordance with human annotations than LDA, especially for themes such as “Multilingual programs,” “Diverse teacher workforce (teacher labor market),” and “Teacher union, salary, and workforce.” Similar patterns were observed in the codewise AUC, which measures a machine’s ability to differentiate true positives from false positives for a theme and is unaffected by data imbalance (see Appendix C Figure~\ref{fig:c6}-\ref{fig:c9}).

\begin{table*}
\centering
\caption{Bootstrapped Performance Metrics}
\resizebox{\textwidth}{!}{
\begin{tabular}{lcccccc}
\toprule
& \multicolumn{3}{c}{Bootstrapped Performance Metrics (Child codes)} & \multicolumn{3}{c}{Bootstrapped Performance Metrics (Parent codes)} \\
\cmidrule(lr){2-4} \cmidrule(lr){5-7}
& Accuracy & Cohen’s $\kappa$ & AUC & Accuracy & Cohen’s $\kappa$ & AUC \\
\midrule
GPT-4 vs. Human & 0.9069  & 0.3738 & 0.7489 & 0.7975 & 0.4570 & 0.7948\\
 &  (95\% CI: 0.9042, 0.9088) &  (95\% CI: 0.3644, 0.3758) &(95\% CI: 0.7383, 0.7596) & (95\% CI: 0.7879, 0.8053) &  (95\% CI: 0.4551, 0.4605) &  (95\% CI: 0.7820, 0.8059) \\
LDA vs. Human & 0.8948  & 0.1862 & 0.6307 & 0.7607 & 0.2928  & 0.6761 \\
 & (95\% CI: 0.8921, 0.8971) & (95\% CI: 0.1850, 0.1899) &  (95\% CI: 0.6200, 0.6407) &(95\% CI: 0.7536, 0.7679) & (95\% CI: 0.2903, 0.2987) & (95\% CI: 0.6606, 0.6878) \\
\bottomrule
\end{tabular}
} 
\label{tab:bootstrapped}
\end{table*}

\begin{figure}[h]
  \centering
  \includegraphics[height=0.45\textheight, keepaspectratio]{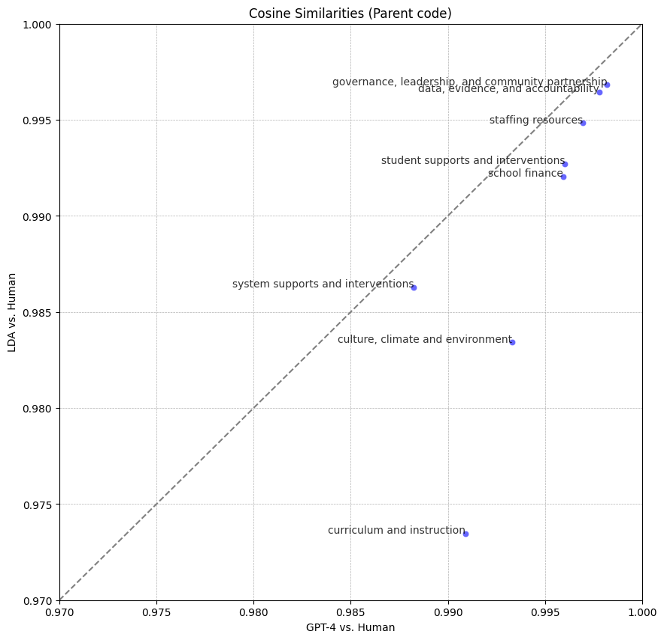}
  \includegraphics[height=0.45\textheight, keepaspectratio]{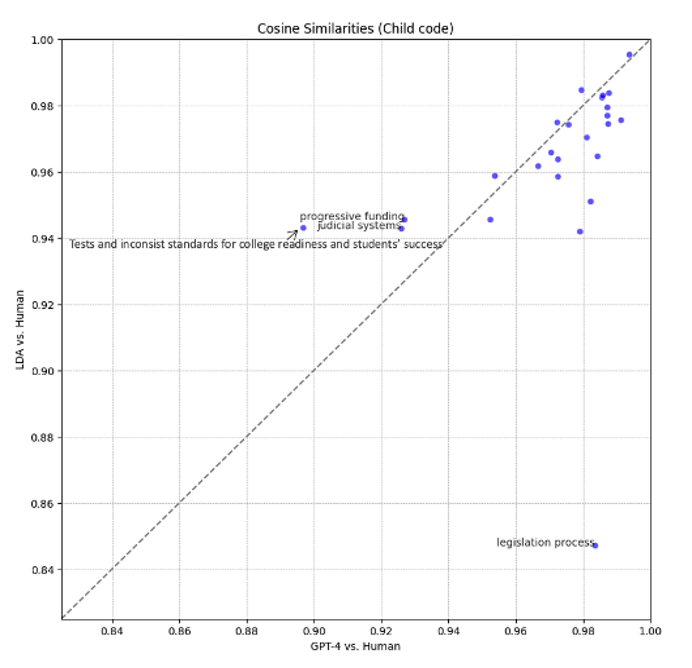}
  \caption{Codewise Cosine Similarity between NLP Thematic Results and Human Coding Results}
  \label{fig:cosine_simil}
  \Description{comparing codewise cosine similarity between LDA vs. human and GPT vs. human}
\end{figure}

We then compared machine and human coding at text level. Figure~\ref{fig:cosine_simil} illustrated a high text similarity for both GPT-4 vs. human and LDA vs. human. Cosine similarities calculated using tf-idf\footnote{tf-idf adjusted for weights for terms to lift the unique terms in a class and downweight universal terms across multiple classes.} are above 0.97 for all parent codes and above 0.94 for 24 child codes (see Appendix C Figure~\ref{fig:c10}-\ref{fig:c12} for zoomed-in figures with code labels). For the majority of child codes and all parent codes, GPT-4 outperformed LDA in this measure, suggesting a closer alignment between GPT-4 and human coding. Notable exceptions included the child codes “Progressive funding,” “Judicials system,” and “Tests and inconsistent standards for college readiness and students’ success,” where GPT-4’s performance was not superior. Moreover, LDA exhibited higher text similarity than GPT-4 for the child code “Data access, analysis, reporting, use, quality, and transparency.” Although “Multilingual programs” achieved the highest agreement between machine and human among all child codes based on both Cohen’s $\kappa$ and the Area Under the Curve (AUC), it did not rank highest for text similarity.

Summarizing all the evaluations, it appears that GPT-4 is capable of identifying themes from context-rich interview data, despite the complexity of the tasks posed by the hierarchical structure of our codebook and the large number of codes. GPT-4 recognized the same themes that were identified by human coders and also picked up on themes that were not selected by human coders. These deviations could potentially enrich the human interpretation of the text, adding nuance to the semantic analysis. However, agreement varied significantly across different themes. In general, GPT-4 performed better at identifying broader themes (parent codes) than more specific themes\footnote{Higher agreement on themes that are less domain specific, including multilingual programs; diversity teacher workforce; teacher union, salary workforce; funding formula; school board; data access, analysis, reporting, use, quality, and transparency.} (child codes), and was more adept at recognizing less domain-specific themes than more domain-specific ones\footnote{Low agreement on themes that are more domain specific themes , including “Progressive funding,” “Local control and district policies and politics,” “Data capacity,” “Trauma at home,” “Instructional programs.”}. Across both child and parent code levels, GPT-4 substantially outperformed LDA on average, although LDA might be more suitable for certain specific themes.

\subsubsection{Qualitative Review for Differences between GPT-4 and Human Thematic Analyses}

\begin{figure}[h]
  \centering
  \includegraphics[width=\linewidth]{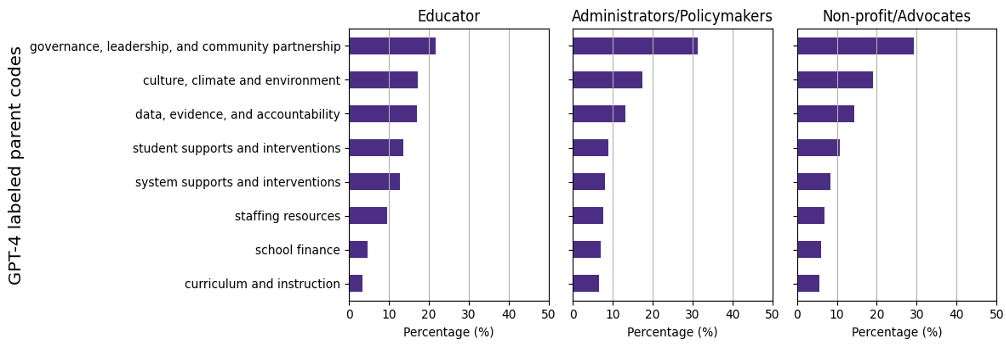}
  \includegraphics[width=\linewidth]{figure6-1.png}
  \caption{GPT-4 Labeled Theme Frequency by Stakeholders’ Job Roles}
  \label{fig:gpt_theme}
  \Description{GPT labeled theme distribution}
\end{figure}

For thematic analysis using GPT-4, we categorized themes by stakeholders’ job roles and summarized the findings. Unlike the uniform patterns of topic distribution among these three types of stakeholders’ job roles revealed by human coding as shown in Figure~\ref{fig:hc-freq}, the GPT-4 results presented in Figure~\ref{fig:gpt_theme} displayed a distinct topic distribution for educators. Specifically, educators focused extensively on themes such as “Anti-racism” and “Trauma at home” within the “Culture, climate, and environment” category, as well as on “Community,” “Student supports,” and “School supports.” These prevalent themes were consistent with our hypothesized focal areas based on educators’ daily responsibilities, which further cast confidence in the utility of GPT-4 coding.

However, the theme “Local control and district policies and politics” was notably absent from educators’ narratives. Considering that the parent category of this child theme remained a significant topic among educators, we propose three potential explanations for this omission. First, the informal expertise that human coders utilized during coding may not have been fully captured in the child code description in the codebook. Second, the definition of this child code, which encompassed district-level decision-making and engagement with partners, as well as accountability and compliance under local control, might have been too nuanced for the LLM to discern, particularly in differentiating it from other child codes under the same parent category. Third, the plethora of codes provided may have overwhelmed GPT-4, preventing it from consistently attending to all codes. For instance, one educator’s statement — “In terms of local policies, something that can be limiting is the emphasis on compliance. I was a teacher in the district, and then I left for several years, and then I came back as a professiona---development specialist around English learners [...]”---clearly fell under this code and its definition but was not accurately labeled by GPT-4.

Moreover, compared to the frequency of human coding, GPT-4 identified the “Culture, climate, and environment” category more frequently. Given that our interviews focused centrally on the evidence for equity in WA K-12 public education, it is not surprising that themes such as “Anti-racism” were prominent in the discussions. GPT-4 seemed to detect nuances that had been overlooked by human coders. For example, one interviewee discussed changes in student demographics in programs for highly capable and gifted students, hinting at increased opportunities for nonwhite-middle-class students. GPT-4 recognized the inherent anti-racism tone in this statement. Nonetheless, GPT-4 also tended to overgeneralize broader themes because it lacked the ability to prioritize themes selectively, as human coders do by choosing fewer than three themes. Additionally, GPT-4 struggled to distinguish between codes with overlapping meanings under the same parent categories, resulting in some instances where themes human coders identified as “Progressive funding” were labeled as “Targeted funding” by GPT-4, inflating the latter’s frequency.

In conclusion, GPT-4 demonstrated proficiency in identifying underlying themes in our interview data. The results also suggest that human coders and LLMs can be complementary, with humans providing the priority, specificity and expertise needed for detailed analysis and judgment, while GPT-4 uncovers embedded meanings that may be overlooked by human analysis.

\subsection{MRQ 2. How accurate and valid are GPT-4 sentiment classifications when comparing to human experts’ and lexicon-based sentiment analysis?}
\subsubsection{Evaluation for Sentiment Analysis}

The confusion matrices in Table~\ref{tab:sentiment_confusion} indicate that GPT-4 aligned more closely with human sentiment labels than the lexical-based VADER. VADER notably overstated the positive sentiment in the text. Although the default settings of VADER, which were used, had been tested and validated by its developers for a variety of contexts, they may not have adapted well to the domain-specific language in educational policy and failed to identify the underlying dissatisfactory sentiments embedded in stakeholders’ descriptions of programs and policies. Evidently, GPT-4 performed better at identifying the sentiment, aligning with human judgment in 58\% of the corpus, especially for “understanding” the expressions of satisfaction, the potential for enhancing equity, and compliments on improvements. However, GPT-4 tended to overestimate sentiment. A significant divergence was observed between human perceptions and GPT-4 classifications when distinguishing “Negative” sentiment from “Neutral,” which impeded the machine’s overall performance. Humans might be more adept at detecting challenges and demands for improvement that were conveyed in the narratives. The inter-rater reliability between human coder and GPT-4 was moderate as measured by Cohen’s $\kappa$ in the last column of Table~\ref{tab:sentiment_confusion}. Cohen’s $\kappa$ for sentiment analysis showed variability in previous studies, with agreement levels differing even among human annotators. For example, Takala et al.[2014]\cite{takala2014gold} reported that some pairwise agreement varied from 0.62 to 0.90 and Cohen’s $\kappa$ varied from 0.41 to 0.80 between experienced human annotators using the three-category sentiments on a common set of economy news stories.

\begin{table*}
\centering
\caption{Lexicon Performance Metrics Comparison}
\resizebox{\textwidth}{!}{
\begin{tabular}{lccccccclcc}
\toprule
& \multicolumn{3}{c}{GPT-4 $\rightarrow$} & \multicolumn{3}{c}{Lexicon $\rightarrow$} & \multicolumn{3}{c}{Performance Metrics} \\
\cmidrule(lr){2-4} \cmidrule(lr){5-7} \cmidrule(lr){8-10}
Human $\downarrow$ & Positive & Negative & Neutral & Positive & Negative & Neutral & & Accuracy & Cohen's $\kappa$ \\
\midrule
Positive & 218 & 4 & 20 & 215 & 11 & 16 & GPT-4 vs.Human & 0.58 & 0.38 \\
Negative & 71 & 322 & 162 & 347 & 151 & 57 & LDA vs. Human & 0.31 & 0.09 \\
Neutral & 31 & 31 & 215 & 405 & 64 & 43 & & & \\
\bottomrule
\end{tabular}
} 
\label{tab:sentiment_confusion}
\end{table*}

\subsubsection{Qualitative Review for Differences between GPT-4 and Human Sentiment Analysis}
\begin{figure}[h]
  \centering
  \includegraphics[width=\linewidth]{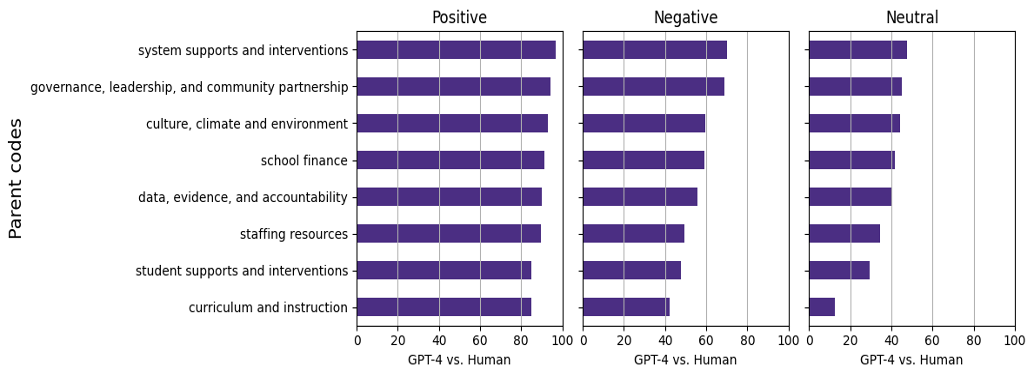}
  \includegraphics[width=\linewidth]{figure7-1.png}
  \caption{Percentage of GPT-4 Sentiment Annotation Agree with Human Annotation by Themes}
  \label{fig:gpt_sentiment}
  \Description{GPT labeled sentiment compared to human coded sentiment}
\end{figure}

Similar to the findings from thematic analysis, the agreement between GPT-4 and humans varied by theme. Figure~\ref{fig:gpt_sentiment} suggests that areas with more direct connections to equity tended to have higher agreement levels, as stakeholders could more clearly articulate their opinions on whether a given practice or policy was equity-enhancing or limiting. The more explicit illustrations resulted in easier sentiment capture by the machine. For example, when mentioning the judicial system, stakeholders often clearly demanded more equitable disciplinary practices. GPT-4 could accurately identify the negative tone in those statements, such as “You will see native kids getting suspended in detention more often than not than their peers. And that comes back to the things that we already talked about. they are seen as problematic or some behavior that they are doing, their peers doing, they do not even blink twice at this brown kid, and like, nope.”

GPT-4 struggled to distinguish mixed feelings, leaning toward neutral rather than negative. For instance, an educator’s statement on learning standards, “I do not even know where I stand, really, on common core standards. I like the idea of standards. I think that does help deliver a more equitable curriculum to students. but I think they are still fuzzy enough that it just simply does not happen. I know it does not happen,” initially acknowledges the positive aspects of learning standards but then y dismisses their effective implementation. In this case, the educator did not indicate satisfaction or improvement over past practices, while highlighting the gap between optimistic intent of the policy and its inefficacious real-world implementation. Consequently, human coders labeled the sentiment as “Negative.” In contrast, GPT-4 recognized the mixed emotions but it was prone to “Neutral” classification because “while they like the idea of standards and believe it can deliver a more equitable curriculum, they also mention that the standards are still fuzzy and not effectively implemented. The statement does not clearly lean towards a positive or negative sentiment, making it neutral.” Although the analyses from human coders and GPT-4 were similar, human coders utilizing their domain knowledge could discern the underlying negative emphasis by understanding the logic in the paragraph. Such expressions of mixed feelings, addressing the deviation from policy intentions and implementation outcomes were more common in the areas like programs and curriculum, funding and accountability, and outcome measures. In these areas, GPT-4 tended to be more optimistic compared to human coders.

One advantage of GPT-4 lies in its approach to analyze each paragraph independently, unlike human coders who were labeling with their preconceived notions about a given educational issue. This indicates the double-edged sword of human domain knowledge or understanding of the background in coding interview texts. In contrast, the detachment of machine from any lived experience or prior knowledge allowed GPT-4 to focus solely on the current paragraph, bringing a fresh perspective that uncovered sentiments in descriptive narratives. In a self-introduction paragraph, an interviewee described his job responsibilities as “I also work a lot with data and digital data systems, so we can get data entered or assumes students digitally when it is appropriate and then the platforms that can bring all that data together so we can use it and integrate it and have it available at our fingertips and make better decisions for student learning.” GPT-4 recognized the satisfaction with the use of data and digital data systems in education and classified this paragraph as a positive expression, while human coders deemed it a neutral description.  

In conclusion, while human coders are better at logical reasoning for differentiating more nuanced sentiment intensity, GPT-4 also has its unique advantage of independent evaluation for each paragraph and uncovering implicit emotions, particularly in descriptive narratives. These distinct capabilities suggest that human and GPT-4 can complement each other in sentiment analysis to combine domain expertise with fresh perspectives. 

\section{Discussion}
In the Washington state K-12 public school system, stakeholders have shown common interests and concerns, while also giving more attention to the areas most relevant to their job roles and responsibilities. LLMs hold great potential for identifying these specific areas along with the associated interests or concerns. This is one of the first studies to examine the potential of the current most advanced LLMs---represented by GPT-4---to facilitate highly domain-specific and context-dependent textual data analysis to facilitate high-stake decision-making. Our dual substance and methodological inquiries in this study have several implications for both educational policy for advancing racial and economic equity, and the potential promises and pitfalls of using LLMs. 

Different stakeholder groups---educators, administrators and policymakers, and non-profit advocates---showed unique thematic focuses in the realm of educational policy. This diversity in thematic emphasis reflects the unique perspectives and priorities of each group. Educators predominantly concentrated on diversifying the teacher workforce and enhancing student supports, underscoring a direct engagement with the educational process. School administrators and policymakers, on the other hand, were more concerned with school finance and legislative processes, indicating a focus on the structural and regulatory aspects of education. Non-profit representatives brought attention to governance and community issues, highlighting the broader social context in which education operates. The varied thematic focuses suggest the need for a multi-faceted approach to educational policy analysis that takes into account the diverse priorities and perspectives of all stakeholders involved. 

Our findings in sentiment analysis show positive sentiments were predominantly directed towards “Progressive funding” and reforms in student support, particularly highlighting the value of “Multilingual programs” in preserving students’ cultural and linguistic heritage. This positive outlook underscores a growing recognition of the importance of cultural inclusivity in education. Conversely, the analysis also identified significant challenges, especially in areas like social-emotional and mental health support, which have been further exacerbated by the COVID-19 pandemic. Another notable concern raised in the analysis is the need for policymakers to prioritize diverse voices in decision-making processes. This includes balancing resource allocation, enhancing inclusivity, and extending community engagement to address the unique needs and perspectives of various groups within the education system. These challenges present barriers to achieving educational equity. For instance, limited resources hinder comprehensive support for students’ diverse needs. The impact of these barriers extends beyond student access to also influence teacher recruitment and retention, particularly for teachers of color. Stakeholders’ sentiments underscore the necessity for more nuanced and responsive educational policies that emphasize the voices and experiences of those most directly involved in the educational process, such as students, teachers, and community members.

Methodologically, GPT-4 demonstrates the large potential of LLMs to assist with analyzing large corpora of data in a timely manner to facilitate domain-specific decision making in educational policy, especially in recognizing broader themes (parent codes). For example, it effectively identified broad themes at the parent code level and the majority of specific themes at the child code level. Besides its excellent performance in identifying themes with clear focal points, like “Multilingual programs” and “Data access, analysis, reporting, and use,” GPT-4 was capable of capturing nuances within broader themes like “Legislation process.” Compared to traditional NLP approaches, such as LDA for thematic analysis and lexical-based sentiment analysis, LLMs have performed exceedingly well in various aspects. LLMs, to some extent, “understand” the nuances of language and domain-specific contexts, which traditional LDA or lexical-based analyses are not equipped to handle. This suggests that GPT-4’s advanced natural language processing capabilities make it well-suited for capturing overarching themes in educational policy discussions.

GPT-4 can complement human expertise in that (a) it can discover nuances overlooked by human coders and (b) keep objective and consistent coding schema without being biased by human’s implicit and unconscious influence from their lived experiences, prior knowledge or cognitive tiredness during the coding. For example, GPT-4 captured nuances in stakeholder discussions about equity related to themes of “Anti-racism” and “Trauma at home,” which were not initially picked up by human coders. Nonetheless, the metric measurements do not fully capture GPT-4’s performance. A meticulous review of GPT-4’s coding output, in comparison with human coders, suggests that the discrepancies between the two sets of results cannot be solely ascribed to GPT-4’s errors, revealing that GPT-4 identified child themes and sentiments often align with the text and added nuances in addition to those selected by human coders. Thus, GPT-4 could effectively complement human coding in both thematic and sentiment analysis. This observation is consistent with prior research; for instance, Liang et al. [2023]\cite{liang2023can} noted that GPT-4’s generated peer review feedback focused on specific aspects of scientific feedback. Moreover, in their survey, over half of the authors whose papers were reviewed by GPT-4 found its feedback helpful or very helpful, and 82.4\% deemed it more beneficial than feedback from some human reviewers.

Additionally, LLMs offer a significant advantage in terms of time efficiency. Human qualitative coding can be labor-intensive; for example, in the initial round using ground theory, three expert coders devoted approximately 20 hours weekly over three months (about 720 hours in total) to complete all coding tasks. This process is also susceptible to the conceptual and subjective biases of the researchers. Collaboration with computational tools markedly reduces the time required for coding. In the subsequent round of human coding with an NLP-informed codebook, two research assistants managed to familiarize themselves with the data and codes and complete the coding in a combined total of 42 hours. Including codebook development, the second stage took approximately 60 hours, a mere fraction of the time spent in the first stage. The use of LLM, specifically GPT-4, further decreased the time required, completing the coding in just 6 hours and introducing new insights into thematic discovery. Therefore, while recognizing the strengths and limitations of LLMs, a synergistic approach between LLMs and human expertise in textual analysis can enhance both efficiency and accuracy.

However, we learned that the performance of LLMs is sensitive to the nature of prompts, varying with domain intensity and depending on the clarity of code descriptions. Prompts that incorporate extensive domain knowledge can help bridge the information gaps evident in current LLMs. The content validity of LLMs in aiding domain-specific data analysis relies on integrating domain knowledge into prompt development, which includes the codebook to be incorporated into the prompt. Although human’s domain knowledge and lived experience may lead to bias and inconsistent implementation of coding schema, they ensure and verify the validity of LLMs and serve as an invaluable form of informal expertise that enriches LLMs’ judgment. 

Moreover, human experts are adept at capturing nuances that LLMs may overlook. GPT-4 showed a tendency to overgeneralize themes, lacking the selective prioritization that human experts exhibit. Notably, GPT-4 performed well in identifying broader themes (parent codes) with an F1 score of 62\%, which is higher than previous studies using similar technologies. However, its performance was less effective at the more specific child code level, struggling to differentiate between closely related sub-themes within the same broader categories. This issue suggests the need for more detailed descriptions in the codebook or prompts to improve LLMs’ precision. Furthermore, GPT-4 might omit information while coding, possibly due to our codebook being overly substantial for its understanding, indicating a potential area for refinement in future research methodologies. We recommend future work to explicitly direct the model to consider all available options before producing results, add more iterations with examples, or conduct multiple analyses on subsets of the data or codebook (e.g., identify the top layer (parent codes) then bottom layer (child codes)). Therefore, human experts remain indispensable throughout the analytical process.

\section{Conclusion}
This study explored the potential of LLMs as a tool for uncovering themes and sentiments embedded in the narratives from stakeholders in Washington K-12 public education. Our study revealed the need for a comprehensive approach to facilitate decision-making that takes into account the diverse priorities and perspectives among stakeholders across different facets of the educational landscape to produce responsive education policies. We conceptualized a resource equity framework, from which we derived a hierarchical structured codebook consisting of 8 parent codes and 28 child codes. Combined with carefully designed prompts, GPT-4 successfully identified the majority of themes (77.89\%) at the child code level that were also identified by human coders and performed exceedingly well at the parent code level, with a 96.02\% overlap with human coders. For sentiment analysis, GPT-4 outperformed the lexical-based method, accurately identifying sentiment in 58\% of the paragraphs. Most importantly, it provided novel perspectives on many texts, which were not identified by human experts. The thematic and sentiment analyses have demonstrated GPT-4’s potential in educational policy studies to analyze stakeholders’ lived experiences and inform policymaking. Despite these promising results, LLMs are not yet capable of performing such analyses independently. Human domain-specific expertise remains crucial throughout the process for guidance and as a quality checker, since the risks such as neglect, difficulty in making fine distinctions, and a tendency for overgeneralization still need to be addressed. Policymakers and researchers should be cognizant of the limitations of LLMs as analytical tools, especially in terms of capturing the specificities of domain-specific meanings. Therefore, a balanced approach that combines the efficient thematic analysis capabilities of LLMs with the nuanced understanding of human coders can lead to more comprehensive and inclusive educational policy analysis that attends the varied needs and priorities of all stakeholders.

\begin{acks}
This study is supported by Baller Group, William T. Grant Foundation (Grant No. 190735), and the National Science Foundation (Grant No. 2055062). Any opinions, findings, and conclusions or recommendations expressed in this material are those of the authors and do not necessarily reflect the views of the funders.
\end{acks}

\bibliographystyle{ACM-Reference-Format}


\begin{thebibliography}{51}


\ifx \showCODEN    \undefined \def \showCODEN     #1{\unskip}     \fi
\ifx \showDOI      \undefined \def \showDOI       #1{#1}\fi
\ifx \showISBNx    \undefined \def \showISBNx     #1{\unskip}     \fi
\ifx \showISBNxiii \undefined \def \showISBNxiii  #1{\unskip}     \fi
\ifx \showISSN     \undefined \def \showISSN      #1{\unskip}     \fi
\ifx \showLCCN     \undefined \def \showLCCN      #1{\unskip}     \fi
\ifx \shownote     \undefined \def \shownote      #1{#1}          \fi
\ifx \showarticletitle \undefined \def \showarticletitle #1{#1}   \fi
\ifx \showURL      \undefined \def \showURL       {\relax}        \fi
\providecommand\bibfield[2]{#2}
\providecommand\bibinfo[2]{#2}
\providecommand\natexlab[1]{#1}
\providecommand\showeprint[2][]{arXiv:#2}

\bibitem[Abdulaziz et~al\mbox{.}(2021)]%
        {abdulaziz2021topic}
\bibfield{author}{\bibinfo{person}{Manal Abdulaziz}, \bibinfo{person}{Alanoud Alotaibi}, \bibinfo{person}{Mashail Alsolamy}, {and} \bibinfo{person}{Abeer Alabbas}.} \bibinfo{year}{2021}\natexlab{}.
\newblock \showarticletitle{Topic based sentiment analysis for COVID-19 tweets}.
\newblock \bibinfo{journal}{\emph{International Journal of Advanced Computer Science and Applications}} \bibinfo{volume}{12}, \bibinfo{number}{1} (\bibinfo{year}{2021}).
\newblock


\bibitem[{Alliance For Resource Equity}(2023)]%
        {dimensions2023}
\bibfield{author}{\bibinfo{person}{{Alliance For Resource Equity}}.} \bibinfo{year}{2023}\natexlab{}.
\newblock \bibinfo{title}{Dimensions of equity}.
\newblock \bibinfo{howpublished}{\url{https://educationresourceequity.org/dimensions-of-equity/}}.
\newblock
\newblock
\shownote{Accessed: 2023-09-14}.


\bibitem[Ashwin et~al\mbox{.}(2023)]%
        {ashwin2023using}
\bibfield{author}{\bibinfo{person}{Julian Ashwin}, \bibinfo{person}{Aditya Chhabra}, {and} \bibinfo{person}{Vijayendra Rao}.} \bibinfo{year}{2023}\natexlab{}.
\newblock \showarticletitle{Using Large Language Models for Qualitative Analysis can Introduce Serious Bias}.
\newblock \bibinfo{journal}{\emph{arXiv preprint arXiv:2309.17147}} (\bibinfo{year}{2023}).
\newblock


\bibitem[Aslett et~al\mbox{.}(2022)]%
        {aslett2020}
\bibfield{author}{\bibinfo{person}{Kevin Aslett}, \bibinfo{person}{Nora {Webb Williams}}, \bibinfo{person}{Andreu Casas}, \bibinfo{person}{Wesley Zuidema}, {and} \bibinfo{person}{John Wilkerson}.} \bibinfo{year}{2022}\natexlab{}.
\newblock \showarticletitle{What Was the Problem in Parkland? Using Social Media to Measure the Effectiveness of Issue Frames}.
\newblock \bibinfo{journal}{\emph{Policy Studies Journal}} \bibinfo{volume}{50}, \bibinfo{number}{1} (\bibinfo{date}{Feb.} \bibinfo{year}{2022}), \bibinfo{pages}{266--289}.
\newblock
\showISSN{0190-292X}
\urldef\tempurl%
\url{https://doi.org/10.1111/psj.12410}
\showDOI{\tempurl}
\newblock
\shownote{Publisher Copyright: {\textcopyright} 2020 Policy Studies Organization}.


\bibitem[Azungah(2018)]%
        {azungah2018qualitative}
\bibfield{author}{\bibinfo{person}{Theophilus Azungah}.} \bibinfo{year}{2018}\natexlab{}.
\newblock \showarticletitle{Qualitative research: deductive and inductive approaches to data analysis}.
\newblock \bibinfo{journal}{\emph{Qualitative research journal}} \bibinfo{volume}{18}, \bibinfo{number}{4} (\bibinfo{year}{2018}), \bibinfo{pages}{383--400}.
\newblock


\bibitem[Baumer et~al\mbox{.}(2017)]%
        {baumer2017comparing}
\bibfield{author}{\bibinfo{person}{Eric~PS Baumer}, \bibinfo{person}{David Mimno}, \bibinfo{person}{Shion Guha}, \bibinfo{person}{Emily Quan}, {and} \bibinfo{person}{Geri~K Gay}.} \bibinfo{year}{2017}\natexlab{}.
\newblock \showarticletitle{Comparing grounded theory and topic modeling: Extreme divergence or unlikely convergence?}
\newblock \bibinfo{journal}{\emph{Journal of the Association for Information Science and Technology}} \bibinfo{volume}{68}, \bibinfo{number}{6} (\bibinfo{year}{2017}), \bibinfo{pages}{1397--1410}.
\newblock


\bibitem[Belal et~al\mbox{.}(2023)]%
        {belal2023leveraging}
\bibfield{author}{\bibinfo{person}{Mohammad Belal}, \bibinfo{person}{James She}, {and} \bibinfo{person}{Simon Wong}.} \bibinfo{year}{2023}\natexlab{}.
\newblock \showarticletitle{Leveraging chatgpt as text annotation tool for sentiment analysis}.
\newblock \bibinfo{journal}{\emph{arXiv preprint arXiv:2306.17177}} (\bibinfo{year}{2023}).
\newblock


\bibitem[Berry and Herrington(2013)]%
        {berry2013tensions}
\bibfield{author}{\bibinfo{person}{Kimberly~Scriven Berry} {and} \bibinfo{person}{Carolyn~D Herrington}.} \bibinfo{year}{2013}\natexlab{}.
\newblock \showarticletitle{Tensions across federalism, localism, and professional autonomy: Social media and stakeholder response to increased accountability}.
\newblock \bibinfo{journal}{\emph{Educational Policy}} \bibinfo{volume}{27}, \bibinfo{number}{2} (\bibinfo{year}{2013}), \bibinfo{pages}{390--409}.
\newblock


\bibitem[Bhattacharya(2017)]%
        {bhattacharya2017fundamentals}
\bibfield{author}{\bibinfo{person}{Kakali Bhattacharya}.} \bibinfo{year}{2017}\natexlab{}.
\newblock \bibinfo{booktitle}{\emph{Fundamentals of qualitative research: A practical guide}}.
\newblock \bibinfo{publisher}{Taylor \& Francis}.
\newblock


\bibitem[Blei et~al\mbox{.}(2003)]%
        {blei2003latent}
\bibfield{author}{\bibinfo{person}{David~M Blei}, \bibinfo{person}{Andrew~Y Ng}, {and} \bibinfo{person}{Michael~I Jordan}.} \bibinfo{year}{2003}\natexlab{}.
\newblock \showarticletitle{Latent dirichlet allocation}.
\newblock \bibinfo{journal}{\emph{Journal of machine Learning research}} \bibinfo{volume}{3}, \bibinfo{number}{Jan} (\bibinfo{year}{2003}), \bibinfo{pages}{993--1022}.
\newblock


\bibitem[Cheng et~al\mbox{.}(2023)]%
        {cheng2023gpt}
\bibfield{author}{\bibinfo{person}{Liying Cheng}, \bibinfo{person}{Xingxuan Li}, {and} \bibinfo{person}{Lidong Bing}.} \bibinfo{year}{2023}\natexlab{}.
\newblock \showarticletitle{Is GPT-4 a Good Data Analyst?}
\newblock \bibinfo{journal}{\emph{arXiv preprint arXiv:2305.15038}} (\bibinfo{year}{2023}).
\newblock


\bibitem[Chew et~al\mbox{.}(2023)]%
        {chew2023llm}
\bibfield{author}{\bibinfo{person}{Robert Chew}, \bibinfo{person}{John Bollenbacher}, \bibinfo{person}{Michael Wenger}, \bibinfo{person}{Jessica Speer}, {and} \bibinfo{person}{Annice Kim}.} \bibinfo{year}{2023}\natexlab{}.
\newblock \showarticletitle{LLM-Assisted Content Analysis: Using Large Language Models to Support Deductive Coding}.
\newblock \bibinfo{journal}{\emph{arXiv preprint arXiv:2306.14924}} (\bibinfo{year}{2023}).
\newblock


\bibitem[Cohen(1960)]%
        {cohen1960coefficient}
\bibfield{author}{\bibinfo{person}{Jacob Cohen}.} \bibinfo{year}{1960}\natexlab{}.
\newblock \showarticletitle{A coefficient of agreement for nominal scales}.
\newblock \bibinfo{journal}{\emph{Educational and psychological measurement}} \bibinfo{volume}{20}, \bibinfo{number}{1} (\bibinfo{year}{1960}), \bibinfo{pages}{37--46}.
\newblock


\bibitem[Dahouda and Joe(2021)]%
        {dahouda2021deep}
\bibfield{author}{\bibinfo{person}{Mwamba~Kasongo Dahouda} {and} \bibinfo{person}{Inwhee Joe}.} \bibinfo{year}{2021}\natexlab{}.
\newblock \showarticletitle{A deep-learned embedding technique for categorical features encoding}.
\newblock \bibinfo{journal}{\emph{IEEE Access}}  \bibinfo{volume}{9} (\bibinfo{year}{2021}), \bibinfo{pages}{114381--114391}.
\newblock


\bibitem[Dai et~al\mbox{.}(2023)]%
        {dai2023llm}
\bibfield{author}{\bibinfo{person}{Shih-Chieh Dai}, \bibinfo{person}{Aiping Xiong}, {and} \bibinfo{person}{Lun-Wei Ku}.} \bibinfo{year}{2023}\natexlab{}.
\newblock \showarticletitle{LLM-in-the-loop: Leveraging Large Language Model for Thematic Analysis}.
\newblock \bibinfo{journal}{\emph{arXiv preprint arXiv:2310.15100}} (\bibinfo{year}{2023}).
\newblock


\bibitem[Das et~al\mbox{.}(2021)]%
        {das2021comparative}
\bibfield{author}{\bibinfo{person}{Nabanita Das}, \bibinfo{person}{Saloni Gupta}, \bibinfo{person}{Srinjoy Das}, \bibinfo{person}{Shuvam Yadav}, \bibinfo{person}{Trishika Subramanian}, {and} \bibinfo{person}{Nairita Sarkar}.} \bibinfo{year}{2021}\natexlab{}.
\newblock \showarticletitle{A Comparative Study of Sentiment Analysis Tools}. In \bibinfo{booktitle}{\emph{2021 International Conference on Innovative Computing, Intelligent Communication and Smart Electrical Systems (ICSES)}}. IEEE, \bibinfo{pages}{1--7}.
\newblock


\bibitem[Davidson et~al\mbox{.}(2022)]%
        {davidson2022improving}
\bibfield{author}{\bibinfo{person}{P. Davidson}, \bibinfo{person}{C. Arndt-Bascle}, \bibinfo{person}{M-G.~de Liedekerke}, {and} \bibinfo{person}{R. Reyes}.} \bibinfo{year}{2022}\natexlab{}.
\newblock \bibinfo{title}{Improving stakeholder engagement and evidence-based policy making}.
\newblock \bibinfo{howpublished}{\url{https://www.theregreview.org/2022/12/07/davidson-improving-stakeholder-engagement/}}.
\newblock
\newblock
\shownote{Accessed: 2023-01-14}.


\bibitem[DiMaggio et~al\mbox{.}(2013)]%
        {dimaggio2013exploiting}
\bibfield{author}{\bibinfo{person}{Paul DiMaggio}, \bibinfo{person}{Manish Nag}, {and} \bibinfo{person}{David Blei}.} \bibinfo{year}{2013}\natexlab{}.
\newblock \showarticletitle{Exploiting affinities between topic modeling and the sociological perspective on culture: Application to newspaper coverage of US government arts funding}.
\newblock \bibinfo{journal}{\emph{Poetics}} \bibinfo{volume}{41}, \bibinfo{number}{6} (\bibinfo{year}{2013}), \bibinfo{pages}{570--606}.
\newblock


\bibitem[Fedorowicz and Aron(2021)]%
        {fedorowicz2021improving}
\bibfield{author}{\bibinfo{person}{Martha Fedorowicz} {and} \bibinfo{person}{Laudan~Y Aron}.} \bibinfo{year}{2021}\natexlab{}.
\newblock \showarticletitle{Improving evidence-based policymaking: A review}.
\newblock \bibinfo{journal}{\emph{Urban Institute: Washington, DC, USA}} (\bibinfo{year}{2021}).
\newblock


\bibitem[Gao et~al\mbox{.}(2023)]%
        {gao2023collabcoder}
\bibfield{author}{\bibinfo{person}{Jie Gao}, \bibinfo{person}{Yuchen Guo}, \bibinfo{person}{Gionnieve Lim}, \bibinfo{person}{Tianqin Zhan}, \bibinfo{person}{Zheng Zhang}, \bibinfo{person}{Toby Jia-Jun Li}, {and} \bibinfo{person}{Simon~Tangi Perrault}.} \bibinfo{year}{2023}\natexlab{}.
\newblock \showarticletitle{CollabCoder: A GPT-Powered Workflow for Collaborative Qualitative Analysis}.
\newblock \bibinfo{journal}{\emph{arXiv preprint arXiv:2304.07366}} (\bibinfo{year}{2023}).
\newblock


\bibitem[Glaser et~al\mbox{.}(1968)]%
        {glaser1968discovery}
\bibfield{author}{\bibinfo{person}{Barney~G Glaser}, \bibinfo{person}{Anselm~L Strauss}, {and} \bibinfo{person}{Elizabeth Strutzel}.} \bibinfo{year}{1968}\natexlab{}.
\newblock \showarticletitle{The discovery of grounded theory; strategies for qualitative research}.
\newblock \bibinfo{journal}{\emph{Nursing research}} \bibinfo{volume}{17}, \bibinfo{number}{4} (\bibinfo{year}{1968}), \bibinfo{pages}{364}.
\newblock


\bibitem[Gosavi(2022)]%
        {gosavi2022transformer}
\bibfield{author}{\bibinfo{person}{Shubham~Ram Gosavi}.} \bibinfo{year}{2022}\natexlab{}.
\newblock \emph{\bibinfo{title}{Transformer based Detection of Sarcasm and it’s Sentiment in Textual Data}}.
\newblock \bibinfo{thesistype}{Ph.\,D. Dissertation}. \bibinfo{school}{Dublin, National College of Ireland}.
\newblock


\bibitem[Grimmer(2013)]%
        {grimmer2013appropriators}
\bibfield{author}{\bibinfo{person}{Justin Grimmer}.} \bibinfo{year}{2013}\natexlab{}.
\newblock \showarticletitle{Appropriators not position takers: The distorting effects of electoral incentives on congressional representation}.
\newblock \bibinfo{journal}{\emph{American Journal of Political Science}} \bibinfo{volume}{57}, \bibinfo{number}{3} (\bibinfo{year}{2013}), \bibinfo{pages}{624--642}.
\newblock


\bibitem[Grimmer and Stewart(2013)]%
        {grimmer2013text}
\bibfield{author}{\bibinfo{person}{Justin Grimmer} {and} \bibinfo{person}{Brandon~M Stewart}.} \bibinfo{year}{2013}\natexlab{}.
\newblock \showarticletitle{Text as data: The promise and pitfalls of automatic content analysis methods for political texts}.
\newblock \bibinfo{journal}{\emph{Political analysis}} \bibinfo{volume}{21}, \bibinfo{number}{3} (\bibinfo{year}{2013}), \bibinfo{pages}{267--297}.
\newblock


\bibitem[Huang et~al\mbox{.}(2023)]%
        {huang2023benchmarking}
\bibfield{author}{\bibinfo{person}{Yixing Huang}, \bibinfo{person}{Ahmed Gomaa}, \bibinfo{person}{Sabine Semrau}, \bibinfo{person}{Marlen Haderlein}, \bibinfo{person}{Sebastian Lettmaier}, \bibinfo{person}{Thomas Weissmann}, \bibinfo{person}{Johanna Grigo}, \bibinfo{person}{Hassen Ben~Tkhayat}, \bibinfo{person}{Benjamin Frey}, \bibinfo{person}{Udo Gaipl}, {et~al\mbox{.}}} \bibinfo{year}{2023}\natexlab{}.
\newblock \showarticletitle{Benchmarking ChatGPT-4 on ACR Radiation Oncology In-Training (TXIT) Exam and Red Journal Gray Zone Cases: Potentials and Challenges for AI-Assisted Medical Education and Decision Making in Radiation Oncology}.
\newblock \bibinfo{journal}{\emph{Available at SSRN 4457218}} (\bibinfo{year}{2023}).
\newblock


\bibitem[Hutto and Gilbert(2014)]%
        {hutto2014vader}
\bibfield{author}{\bibinfo{person}{Clayton Hutto} {and} \bibinfo{person}{Eric Gilbert}.} \bibinfo{year}{2014}\natexlab{}.
\newblock \showarticletitle{Vader: A parsimonious rule-based model for sentiment analysis of social media text}. In \bibinfo{booktitle}{\emph{Proceedings of the international AAAI conference on web and social media}}, Vol.~\bibinfo{volume}{8}. \bibinfo{pages}{216--225}.
\newblock


\bibitem[Jelodar et~al\mbox{.}(2020)]%
        {jelodar2020deep}
\bibfield{author}{\bibinfo{person}{Hamed Jelodar}, \bibinfo{person}{Yongli Wang}, \bibinfo{person}{Rita Orji}, {and} \bibinfo{person}{Shucheng Huang}.} \bibinfo{year}{2020}\natexlab{}.
\newblock \showarticletitle{Deep sentiment classification and topic discovery on novel coronavirus or COVID-19 online discussions: NLP using LSTM recurrent neural network approach}.
\newblock \bibinfo{journal}{\emph{IEEE Journal of Biomedical and Health Informatics}} \bibinfo{volume}{24}, \bibinfo{number}{10} (\bibinfo{year}{2020}), \bibinfo{pages}{2733--2742}.
\newblock


\bibitem[Kheiri and Karimi(2023)]%
        {kheiri2023sentimentgpt}
\bibfield{author}{\bibinfo{person}{Kiana Kheiri} {and} \bibinfo{person}{Hamid Karimi}.} \bibinfo{year}{2023}\natexlab{}.
\newblock \showarticletitle{Sentimentgpt: Exploiting gpt for advanced sentiment analysis and its departure from current machine learning}.
\newblock \bibinfo{journal}{\emph{arXiv preprint arXiv:2307.10234}} (\bibinfo{year}{2023}).
\newblock


\bibitem[Kojima et~al\mbox{.}(2022)]%
        {kojima2022large}
\bibfield{author}{\bibinfo{person}{Takeshi Kojima}, \bibinfo{person}{Shixiang~Shane Gu}, \bibinfo{person}{Machel Reid}, \bibinfo{person}{Yutaka Matsuo}, {and} \bibinfo{person}{Yusuke Iwasawa}.} \bibinfo{year}{2022}\natexlab{}.
\newblock \showarticletitle{Large language models are zero-shot reasoners}.
\newblock \bibinfo{journal}{\emph{Advances in neural information processing systems}}  \bibinfo{volume}{35} (\bibinfo{year}{2022}), \bibinfo{pages}{22199--22213}.
\newblock


\bibitem[Leeson et~al\mbox{.}(2019)]%
        {leeson2019natural}
\bibfield{author}{\bibinfo{person}{William Leeson}, \bibinfo{person}{Adam Resnick}, \bibinfo{person}{Daniel Alexander}, {and} \bibinfo{person}{John Rovers}.} \bibinfo{year}{2019}\natexlab{}.
\newblock \showarticletitle{Natural language processing (NLP) in qualitative public health research: a proof of concept study}.
\newblock \bibinfo{journal}{\emph{International Journal of Qualitative Methods}}  \bibinfo{volume}{18} (\bibinfo{year}{2019}), \bibinfo{pages}{1609406919887021}.
\newblock


\bibitem[Liang et~al\mbox{.}(2023)]%
        {liang2023can}
\bibfield{author}{\bibinfo{person}{Weixin Liang}, \bibinfo{person}{Yuhui Zhang}, \bibinfo{person}{Hancheng Cao}, \bibinfo{person}{Binglu Wang}, \bibinfo{person}{Daisy Ding}, \bibinfo{person}{Xinyu Yang}, \bibinfo{person}{Kailas Vodrahalli}, \bibinfo{person}{Siyu He}, \bibinfo{person}{Daniel Smith}, \bibinfo{person}{Yian Yin}, {et~al\mbox{.}}} \bibinfo{year}{2023}\natexlab{}.
\newblock \showarticletitle{Can large language models provide useful feedback on research papers? A large-scale empirical analysis}.
\newblock \bibinfo{journal}{\emph{arXiv preprint arXiv:2310.01783}} (\bibinfo{year}{2023}).
\newblock


\bibitem[Liu et~al\mbox{.}(2021)]%
        {liu2021p}
\bibfield{author}{\bibinfo{person}{Xiao Liu}, \bibinfo{person}{Kaixuan Ji}, \bibinfo{person}{Yicheng Fu}, \bibinfo{person}{Weng~Lam Tam}, \bibinfo{person}{Zhengxiao Du}, \bibinfo{person}{Zhilin Yang}, {and} \bibinfo{person}{Jie Tang}.} \bibinfo{year}{2021}\natexlab{}.
\newblock \showarticletitle{P-tuning v2: Prompt tuning can be comparable to fine-tuning universally across scales and tasks}.
\newblock \bibinfo{journal}{\emph{arXiv preprint arXiv:2110.07602}} (\bibinfo{year}{2021}).
\newblock


\bibitem[Lyu et~al\mbox{.}(2023)]%
        {lyu2023translating}
\bibfield{author}{\bibinfo{person}{Qing Lyu}, \bibinfo{person}{Josh Tan}, \bibinfo{person}{Mike~E Zapadka}, \bibinfo{person}{Janardhana Ponnatapuram}, \bibinfo{person}{Chuang Niu}, \bibinfo{person}{Ge Wang}, {and} \bibinfo{person}{Christopher~T Whitlow}.} \bibinfo{year}{2023}\natexlab{}.
\newblock \showarticletitle{Translating radiology reports into plain language using chatgpt and gpt-4 with prompt learning: Promising results, limitations, and potential}.
\newblock \bibinfo{journal}{\emph{arXiv preprint arXiv:2303.09038}} (\bibinfo{year}{2023}).
\newblock


\bibitem[Maxwell(2004)]%
        {maxwell2004causal}
\bibfield{author}{\bibinfo{person}{Joseph~A Maxwell}.} \bibinfo{year}{2004}\natexlab{}.
\newblock \showarticletitle{Causal explanation, qualitative research, and scientific inquiry in education}.
\newblock \bibinfo{journal}{\emph{Educational researcher}} \bibinfo{volume}{33}, \bibinfo{number}{2} (\bibinfo{year}{2004}), \bibinfo{pages}{3--11}.
\newblock


\bibitem[Nguyen et~al\mbox{.}(2021)]%
        {nguyen2021sarcasm}
\bibfield{author}{\bibinfo{person}{Hieu Nguyen}, \bibinfo{person}{Jihye Moon}, \bibinfo{person}{Nijhum Paul}, {and} \bibinfo{person}{Swapna~S Gokhale}.} \bibinfo{year}{2021}\natexlab{}.
\newblock \showarticletitle{Sarcasm detection in politically motivated social media content}. In \bibinfo{booktitle}{\emph{2021 IEEE Intl Conf on Parallel \& Distributed Processing with Applications, Big Data \& Cloud Computing, Sustainable Computing \& Communications, Social Computing \& Networking (ISPA/BDCloud/SocialCom/SustainCom)}}. IEEE, \bibinfo{pages}{1538--1545}.
\newblock


\bibitem[{OpenAI Community}(2023)]%
        {openai2023cheatsheet}
\bibfield{author}{\bibinfo{person}{{OpenAI Community}}.} \bibinfo{year}{2023}\natexlab{}.
\newblock \bibinfo{title}{Cheat Sheet: Mastering Temperature and Top\_p in ChatGPT API}.
\newblock \bibinfo{howpublished}{\url{https://openai.com/developer-forum/cheat-sheet-mastering-temperature-and-top-p-in-chatgpt-api}}.
\newblock
\newblock
\shownote{Retrieved from OpenAI Developer Forum (a few tips and tricks on controlling the creativity/deterministic output of prompt responses.)}.


\bibitem[Patton(1990)]%
        {patton1990qualitative}
\bibfield{author}{\bibinfo{person}{Michael~Quinn Patton}.} \bibinfo{year}{1990}\natexlab{}.
\newblock \bibinfo{booktitle}{\emph{Qualitative evaluation and research methods}}.
\newblock \bibinfo{publisher}{SAGE Publications, inc}.
\newblock


\bibitem[Roberts et~al\mbox{.}(2014)]%
        {roberts2014structural}
\bibfield{author}{\bibinfo{person}{Margaret~E Roberts}, \bibinfo{person}{Brandon~M Stewart}, \bibinfo{person}{Dustin Tingley}, \bibinfo{person}{Christopher Lucas}, \bibinfo{person}{Jetson Leder-Luis}, \bibinfo{person}{Shana~Kushner Gadarian}, \bibinfo{person}{Bethany Albertson}, {and} \bibinfo{person}{David~G Rand}.} \bibinfo{year}{2014}\natexlab{}.
\newblock \showarticletitle{Structural topic models for open-ended survey responses}.
\newblock \bibinfo{journal}{\emph{American journal of political science}} \bibinfo{volume}{58}, \bibinfo{number}{4} (\bibinfo{year}{2014}), \bibinfo{pages}{1064--1082}.
\newblock


\bibitem[Rosenberg et~al\mbox{.}(2021)]%
        {rosenberg2021understanding}
\bibfield{author}{\bibinfo{person}{Joshua~M Rosenberg}, \bibinfo{person}{Conrad Borchers}, \bibinfo{person}{Elizabeth~B Dyer}, \bibinfo{person}{Daniel Anderson}, {and} \bibinfo{person}{Christian Fischer}.} \bibinfo{year}{2021}\natexlab{}.
\newblock \showarticletitle{Understanding public sentiment about educational reforms: The next generation science standards on Twitter}.
\newblock \bibinfo{journal}{\emph{AERA open}}  \bibinfo{volume}{7} (\bibinfo{year}{2021}), \bibinfo{pages}{23328584211024261}.
\newblock


\bibitem[Savelka(2023)]%
        {savelka2023unlocking}
\bibfield{author}{\bibinfo{person}{Jaromir Savelka}.} \bibinfo{year}{2023}\natexlab{}.
\newblock \showarticletitle{Unlocking practical applications in legal domain: Evaluation of gpt for zero-shot semantic annotation of legal texts}.
\newblock \bibinfo{journal}{\emph{arXiv preprint arXiv:2305.04417}} (\bibinfo{year}{2023}).
\newblock


\bibitem[Shlezinger et~al\mbox{.}(2023)]%
        {shlezinger2023model}
\bibfield{author}{\bibinfo{person}{Nir Shlezinger}, \bibinfo{person}{Jay Whang}, \bibinfo{person}{Yonina~C Eldar}, {and} \bibinfo{person}{Alexandros~G Dimakis}.} \bibinfo{year}{2023}\natexlab{}.
\newblock \showarticletitle{Model-based deep learning}.
\newblock \bibinfo{journal}{\emph{Proc. IEEE}} (\bibinfo{year}{2023}).
\newblock


\bibitem[Sprenkamp et~al\mbox{.}(2023)]%
        {sprenkamp2023large}
\bibfield{author}{\bibinfo{person}{Kilian Sprenkamp}, \bibinfo{person}{Daniel~Gordon Jones}, {and} \bibinfo{person}{Liudmila Zavolokina}.} \bibinfo{year}{2023}\natexlab{}.
\newblock \showarticletitle{Large Language Models for Propaganda Detection}.
\newblock \bibinfo{journal}{\emph{arXiv preprint arXiv:2310.06422}} (\bibinfo{year}{2023}).
\newblock


\bibitem[Sun et~al\mbox{.}(2019)]%
        {sun2019using}
\bibfield{author}{\bibinfo{person}{Min Sun}, \bibinfo{person}{Jing Liu}, \bibinfo{person}{Junmeng Zhu}, {and} \bibinfo{person}{Zachary LeClair}.} \bibinfo{year}{2019}\natexlab{}.
\newblock \showarticletitle{Using a text-as-data approach to understand reform processes: A deep exploration of school improvement strategies}.
\newblock \bibinfo{journal}{\emph{Educational Evaluation and Policy Analysis}} \bibinfo{volume}{41}, \bibinfo{number}{4} (\bibinfo{year}{2019}), \bibinfo{pages}{510--536}.
\newblock


\bibitem[Takala et~al\mbox{.}(2014)]%
        {takala2014gold}
\bibfield{author}{\bibinfo{person}{Pyry Takala}, \bibinfo{person}{Pekka Malo}, \bibinfo{person}{Ankur Sinha}, {and} \bibinfo{person}{Oskar Ahlgren}.} \bibinfo{year}{2014}\natexlab{}.
\newblock \showarticletitle{Gold-standard for Topic-specific Sentiment Analysis of Economic Texts.}. In \bibinfo{booktitle}{\emph{LREC}}, Vol.~\bibinfo{volume}{2014}. \bibinfo{pages}{2152--2157}.
\newblock


\bibitem[Tang et~al\mbox{.}(2023)]%
        {tang2023science}
\bibfield{author}{\bibinfo{person}{Ruixiang Tang}, \bibinfo{person}{Yu-Neng Chuang}, {and} \bibinfo{person}{Xia Hu}.} \bibinfo{year}{2023}\natexlab{}.
\newblock \showarticletitle{The science of detecting llm-generated texts}.
\newblock \bibinfo{journal}{\emph{arXiv preprint arXiv:2303.07205}} (\bibinfo{year}{2023}).
\newblock


\bibitem[Wallner(2008)]%
        {wallner2008legitimacy}
\bibfield{author}{\bibinfo{person}{Jennifer Wallner}.} \bibinfo{year}{2008}\natexlab{}.
\newblock \showarticletitle{Legitimacy and public policy: Seeing beyond effectiveness, efficiency, and performance}.
\newblock \bibinfo{journal}{\emph{Policy Studies Journal}} \bibinfo{volume}{36}, \bibinfo{number}{3} (\bibinfo{year}{2008}), \bibinfo{pages}{421--443}.
\newblock


\bibitem[Wang et~al\mbox{.}(2023)]%
        {wang2023chatgpt}
\bibfield{author}{\bibinfo{person}{Jiaan Wang}, \bibinfo{person}{Yunlong Liang}, \bibinfo{person}{Fandong Meng}, \bibinfo{person}{Haoxiang Shi}, \bibinfo{person}{Zhixu Li}, \bibinfo{person}{Jinan Xu}, \bibinfo{person}{Jianfeng Qu}, {and} \bibinfo{person}{Jie Zhou}.} \bibinfo{year}{2023}\natexlab{}.
\newblock \showarticletitle{Is chatgpt a good nlg evaluator? a preliminary study}.
\newblock \bibinfo{journal}{\emph{arXiv preprint arXiv:2303.04048}} (\bibinfo{year}{2023}).
\newblock


\bibitem[Wei et~al\mbox{.}(2022)]%
        {wei2022chain}
\bibfield{author}{\bibinfo{person}{Jason Wei}, \bibinfo{person}{Xuezhi Wang}, \bibinfo{person}{Dale Schuurmans}, \bibinfo{person}{Maarten Bosma}, \bibinfo{person}{Fei Xia}, \bibinfo{person}{Ed Chi}, \bibinfo{person}{Quoc~V Le}, \bibinfo{person}{Denny Zhou}, {et~al\mbox{.}}} \bibinfo{year}{2022}\natexlab{}.
\newblock \showarticletitle{Chain-of-thought prompting elicits reasoning in large language models}.
\newblock \bibinfo{journal}{\emph{Advances in Neural Information Processing Systems}}  \bibinfo{volume}{35} (\bibinfo{year}{2022}), \bibinfo{pages}{24824--24837}.
\newblock


\bibitem[Xiao et~al\mbox{.}(2023)]%
        {xiao2023supporting}
\bibfield{author}{\bibinfo{person}{Ziang Xiao}, \bibinfo{person}{Xingdi Yuan}, \bibinfo{person}{Q~Vera Liao}, \bibinfo{person}{Rania Abdelghani}, {and} \bibinfo{person}{Pierre-Yves Oudeyer}.} \bibinfo{year}{2023}\natexlab{}.
\newblock \showarticletitle{Supporting Qualitative Analysis with Large Language Models: Combining Codebook with GPT-3 for Deductive Coding}. In \bibinfo{booktitle}{\emph{Companion Proceedings of the 28th International Conference on Intelligent User Interfaces}}. \bibinfo{pages}{75--78}.
\newblock


\bibitem[Zhao et~al\mbox{.}(2021)]%
        {zhao2021calibrate}
\bibfield{author}{\bibinfo{person}{Zihao Zhao}, \bibinfo{person}{Eric Wallace}, \bibinfo{person}{Shi Feng}, \bibinfo{person}{Dan Klein}, {and} \bibinfo{person}{Sameer Singh}.} \bibinfo{year}{2021}\natexlab{}.
\newblock \showarticletitle{Calibrate before use: Improving few-shot performance of language models}. In \bibinfo{booktitle}{\emph{International Conference on Machine Learning}}. PMLR, \bibinfo{pages}{12697--12706}.
\newblock


\bibitem[Zhou et~al\mbox{.}(2023)]%
        {zhou2023guided}
\bibfield{author}{\bibinfo{person}{Sulong Zhou}, \bibinfo{person}{Pengyu Kan}, \bibinfo{person}{Qunying Huang}, {and} \bibinfo{person}{Janet Silbernagel}.} \bibinfo{year}{2023}\natexlab{}.
\newblock \showarticletitle{A guided latent Dirichlet allocation approach to investigate real-time latent topics of Twitter data during Hurricane Laura}.
\newblock \bibinfo{journal}{\emph{Journal of Information Science}} \bibinfo{volume}{49}, \bibinfo{number}{2} (\bibinfo{year}{2023}), \bibinfo{pages}{465--479}.
\newblock


\end{thebibliography}

\appendix

\section{Codebook Establish}

\subsection{Interview Protocol (45 minutes)}
Background Questions (5 mins)	

1. Could you please briefly describe your role, your primary responsibilities, and how long you have been in your current role?

2. How do you define racial and economic equity in education? What are the specific equity-related goals you/your organization are striving to achieve?
\\\\	
Current Policies (15 mins) 

[We will need to do some background study of inter- viewee’s work before the interview]

3. From your work and based on your perspective, can you offer 1 or 2 examples of current state and/or local policies that you think most enhance racial and economic equity in the Washington State public education system?
					
[Probe] Why and how do you think this policy/strategy has enhanced racial and eco- nomic equity?

[If you sense the interviewees have expertise to comment on any of the following, please probe:]	

\begin{itemize}
    \item (3.1) In terms of school finance, can you think of ways that state and local policies enhance racial and economic equity?
    \item (3.2) In terms of school governance, can you think of ways that state and local decision-making structures and processes enhance equity?
    \item (3.3). In terms of teacher resources, can you think of ways that current policies promote racial and economic equity?	
\end{itemize}
	 					
4. From your work and based on your perspective, can you offer 1 or 2 examples of current state and/or local policies that you think most limit racial and economic equity in the Washington State public education system?	

[Probe] Why and how do you think this policy/strategy has limited racial and eco- nomic equity?		

[If you sense the interviewees have expertise to comment on any of the following, please probe:]	

\begin{itemize}
    \item (4.1) In terms of school finance, what factors hinder the equitable distribution of resources across groups of students based on race and income? How could the state more equitably distribute resources across geographically regions?
    \item (4.2) In terms of school governance, how do current policies or processes hinder collaboration to promote equity?\\The relationship between the teacher union and school district is also critical for school governance. Could you also comment on how to build a better rela- tionship between these two parties?
    \item (4.3). In terms of teacher resources, do you experience teacher shortage? in what aspects (e.g., ELL, special edu, STEM)? \\What are the major barriers for recruiting and retaining effective teachers, particularly teachers of color?
\end{itemize}

5. We know that COVID-19 has brought devastation to many communities. But we also know that changes of this magnitude present opportunities. Are there any op- portunities to advance racial and economic equity that you see as we “build back” from COVID-19?
\\\\
Information Gaps (15 mins) 
[This next set of questions focuses on current information gaps in the state.]	
		
6. In the context of your work, do you feel that you have access to reliable information or data that you need to effectively do your job?	

[if yes]

-	Which resource/s do you consult most frequently?

-	What do you find most helpful about that resource/those resources?			

[if no]

-	What kind of information could better support policy development/implementation?	

[if you sense the interviewees have expertise to comment on any of the following, please probe:]	

\begin{itemize}
    \item (6.1) Are there specific knowledge gaps around school finance that are limiting actions that can advance racial and economic equity? 
    \item (6.2) Are there specific knowledge gaps around school governance that are limiting actions that can advance racial and economic equity?
    \item (6.3) Are there specific knowledge gaps around teacher distribution/shortages that are limiting actions that can advance racial and economic equity? 
\end{itemize}

\noindent Potential Improvements (15 mins).	

7. In the context of your work, can you think of any specific information, data or tools that could help you make better informed decisions?		

[probe] in your view, who in the state is in the best position to take lead in generating and distributing that information/data or developing that tool?

\subsection{Rubrics and Procedures for Rating Topic Coherence	}
Step 1: Preparation. Using the STM visualization graph, select at least 20 tasks that have the highest proportion loaded on one given topic. Start with the highest proportional ones, then read through the statement of tasks, synthesize key ideas across the tasks, and label each topic.
\\\\				
Step 2: Use the last column in the attached template in the spreadsheet to record your summary (or 2–3 key words) of the exemplary document you are reading. This helps you to clearly apply the rubrics below and keep track of documents you have reviewed.
\\\\				
Step 3: Rate topic coherence using the following metrics for the extent to which the topic is coherent.
\\\\
\fbox{
    \parbox{0.95\textwidth}{
On a 4-point scale: 4 = a great deal; 3 = moderate; 2 = little; 1 = none
\begin{itemize}
    \item 4 = a great deal: It is easy for me to find one coherent latent construct for this topic. The label emerges from the exemplary documents coherently.
    \item 3 = moderate: The topic contains 2–3 latent constructs; however, they are closely related. I am still able to come up with one label to summarize almost all exemplary documents.
    \item 2 = little: The topic contains more than 2 latent constructs that are somewhat connected. I manage to come up with a label, but it only summarizes a portion of the exemplary texts well.
    \item 1 = none: The documents under this topic are largely random, with no clear relationships.
\end{itemize}}}		
\\\\
Step 4: Record your rationales for (a) the label you have created; and (b) the topic coherence rating you have given. 						
			
\subsection{Final Version of Codebook with Agreement Ratings}

\begin{longtable}{p{0.1\linewidth}p{0.15\linewidth}p{0.15\linewidth}p{0.2\linewidth}p{0.1\linewidth}p{0.2\linewidth}}
\caption{Final Version of Codebook with Agreement Ratings} \\
\toprule
Topic \# from LDA & Parent & Child & Child\_description & Agreed rating & Key words \\
\midrule
\endfirsthead
\multicolumn{6}{c}%
{\tablename\ \thetable\ -- \textit{Continued from previous page}} \\
\toprule
Topic \# from LDA & Parent & Child & Child\_description & Agreed rating & Key words \\
\midrule
\endhead
\hline \multicolumn{6}{r}{\textit{Continued on next page}} \\
\endfoot
\bottomrule
\endlastfoot

23 & Culture, climate and environment & Trauma at home & Struggling home and family experiences of children of high poverty and of color negatively influence their school learning and graduation pathways post pandemic & 4 & bad, kid, children, home, school, family, grade, covid-19, hispanic, pandemic, third, rate, happen, stop, number, class, get, year, somebody, want \\
14 & Culture, climate and environment & Anti-racism & Talking about whiteness, success, and anti-racism & 3 & pull, white, child, success, stay, built, job, indic, keep, whole, four, barrier, story, middle, racial, ago, pretty, feel, understand, day \\
None & Curriculum and instruction & Instructional programs & AP/IB courses, college classes/credits in high school, special education programs, bilingual programs, ethnic studies &  & Equity, Students, School, District, Policy, Data, Programs, Honors, Course, AP, IB, State, Access, Advanced, Highly Capable, College, Racial, Learning, Services, Practices. \\
13 & Curriculum and instruction & Curriculum development and instructional delivery & School curriculum development and instructional delivery, specially including culturally responsive teaching and equitable pedagogical practices that influence students' experience in the classroom & 3 & online, teach, taught, teacher, high, learn, class, middle, science, elementary, experience, student, day, next, sit, first, nothing, life, school, potential \\
11 & Data, evidence, and accountability & Data access, analysis, reporting, use, quality and transparency & Data collection, access, analysis, and use to help practitioners improve their practices/strategies and to help policymakers (at all levels - e.g., school building, district, state legislators) make new policies or refine current policies. This also pertains to data sharing and reporting, as well as data transparency and quality issues. & 4 & dashboard, data, information, assessment, collect, disaggregate, report, website, access, use, ospi, yes, effective, analysis, tool, together, good, story, wsif, point \\
19 & Data, evidence, and accountability & Goals, outcomes, and measures: Tests, standards, graduation requirements & Defining education and school improvement goals and measuring outcomes via tests, learning standards, college readiness, and graduation requirements, and preparation for jobs, as well as linking goals with outcome measures. & 3.5 & kind, cours, goal, access, outcom, sure, citi, drive, kid, take, pathway, possible, easi, deeper, make, job, report, honor, pasco, land \\
26 & Data, evidence, and accountability & Tests and inconsistent standards for college readiness and students' success & Tests are inconsistent standards for what the education system should produce, nor are good measures of college readiness. & 3.5 & test, standard, score, tribal, take, math, assumess, let, measure, rate, indic, enroll, consult, college, science, differ, whether, nativ, graduate, ethnic \\
None & Data, evidence, and accountability & Accountability system & Systems used to hold an individual, group or organization responsible for doing something that they are supposed to be doing according to a law, job description or other agreement. Rewards or sanctions are associated with performance. &  & accountability, system, policy, district, outcomes, student, compliance, state, school, data, community, work, focus, measure, governance, metrics, assessment, scores, education, partnership \\
None & Data, evidence, and accountability & Data capacity & Related to capacity of individuals, groups or organizations to process data and information and conduct research of any kind for any purpose &  & data, resources, state, access, learning, research, organizations, agencies, capacity, digital, analysis, trends, support, universities, questions, information, community, school, finance, equity \\
3 & Governance, leadership, and community partnership & Local control and district policies and politics & District decision making, accountability and compliance, under local control, and district engagement with community and parents & 4 & budget, school-board, decis, sometime, make, stakeholder, member, director, engag, power, superintend, process, polit, chang, allow, piec, toward, along, act, district \\
30 & Governance, leadership, and community partnership & Legislation process & Bills and legislative process for educational policies & 3.5 & bill, pass, half, one, read, last, educ, way, year, hous, legisl, make, start, actual, basic \\
28 & Governance, leadership, and community partnership & Leadership in diversity & Leadership, diverse community, and leadership in diversity; can include representation on racial/ethnic and other dimension (e.g., gender identity, disability, immigration status, etc); can also lead to achieve diversity and inclusion & 3 & add, support, divers, nativ, trust, leadership, indian, broad, community, provide, locat, hispan, critic, recogn, place, huge, coalit, role, continu, partner \\
9 & Governance, leadership, and community partnership & Coalition and relationship & Go beyond superficial things (counting and disaggregating numbers) and build relationships and coalition centering the voices of youth and families in marginalized communities & 3 & work, try, hard, native, many, educ, push, tell, center, people, excited, voice, talk, family, count \\
17 & Governance, leadership, and community partnership & School board & School boards' structure, diversity and representation, and accountability system and inquiry cycles. School board governance powers, school board members as leaders, school board elections, expertise (or lack thereof) & 4 & board, plan, perspective, staff, answer, identify, action, guess, educ, suppose, one, experience, idea, believe, cycle \\
20 & Governance, leadership, and community partnership & Community & Factors and strategies that influence the collaboration and building of meaningful connections with communities: leadership capacity, leaders listening to others, seeking advice and feedback from external/served individuals and organizations, working directly with families and bringing them actively into the conversation. & 4 & capacity, term, agency, build, community, depart, collaborate, limit, engage, within, understand, connect, listen, effect, show, table, expect, factor, response, way \\
2 & Governance, leadership, and community partnership & Government relationships & State and local public school systems relationships (decentralization and local control) and relationships with non-profit organizations & 3.5 & positive, question, level, district, local, response, come, different, engage, ask, esd, foundation, begin, forward, study, ospi, try, dynamic, exact, versus \\
25 & School finance & Progressive funding & Special education, federal and state funding for districts with low-income, ELL, special-ed, and multicultural students who need more supports & 4 & special-ed, student, federal, fund, dollar, district, mention, challenge, additional, population, identify, high, interest, versus, ell, flexible, need, learn, improve, school \\
8 & School finance & Funding formula & Funding formula based on districts' needs, local levies, and the allocations of McCleary and Stimulus funds & 4 & formula, money, fund, spend, finance, district, dollar, levy, school, essa, amount, distribute, mccleary, certain, account, tax, financial, receive, away, use \\
18 & School finance & Targeted funds & Targeted funds for specific activities/programs: Funds or grants available for other purposes than teacher salary, funding for teacher preparation and professional development, and targeted investment in community of color & 3 & paid, pay, grant, fund, available, tax, seattle, whether, another, obvious, black, spend, one, teacher, brown, purpose, influence, sps, least, othello \\
10 & Staffing resources & Teacher union, salary, workforce & Teacher union, politics and local collective bargaining processes, as well as teacher unions' influences on teacher salaries, professional development, and teacher hiring & 4 & union, teacher, salary, principle, hire, administration, contract, quality, local, survey, control, value, district, someone, happen, professional-develop, true, year, experience, become \\
24 & Staffing resources & Diversify teacher workforce (teacher labor market) & Teacher education, diversifying teacher candidate pool/pathways, and partnerships between K-12 and higher edu, and with outside organizations & 3 & field, change, part, pathway, teacher-educ, local, force, discuss, invest, partnership, important, structure, k-12, become, nation, outside, impact, prepare, shift, parent \\
16 & Staffing resources & Mentoring, coaching, and teacher learning & New teachers' coaching, mentoring, and PD in equitable instruction and curriculum (adopt, professional learning) & 4 & coach, new, classroom, instruct, role, teacher, practice, equity, culture, term, sure, curriculum, converse, adopt, science, critical, support, colleague, administration, person \\
21 & Student supports and interventions & Differentiated student strategies & Differential and targeted strategies for students of color, particularly American African, Southeast Asian, and Hispanic & 4 & african, group, student, include, american, focus, achieve, strategy, target, race, specific, goal, students--color, kind, important, outcome, system, define, measure, name \\
22 & Student supports and interventions & Multilingual programs & Access and quality of bilingual, multilingual programs offered to English language learners & 4 & dual, language, program, english, bilingual, learner, student, year, service, ell, skill, research, access, million, last, develop, director, past, provide, initiative \\
4 & Student supports and interventions & Students' SEL and health & Resources supporting Social emotional learning (SEL) and youth health & 4 & mental, health, resource, young, space, assist, person, social, large, people, academy, best, care, beyond, create, potential, necessary, challenge, k-12, impact \\
27 & Student supports and interventions & Learning opportunities and programs & Learning opportunities in schools, and tribal and native education & 3 & leave, opportunity, year, different, five, native, tribe, type, office, back, time, last, far, legislate, whole, curriculum, open, everybody, longer, earlier \\
6 & System supports and interventions & School system support and improvement & Tiered support for system and school improvement & 4 & tier, school, high, three, improve, two, year, score, attend, survey, foundation, wsif, elementary, director, perform, district, esd, five, track, metric \\
7 & System supports and interventions & Judicial systems & Court, state, judicial systems, and institution's role in educational equity, particularly racial equity & 4 & court, state, washington, system, committee, county, gap, educ, covid-19, represent, justice, institute, executive, house, govern, guy, thought, children, goes, throughout \\
\end{longtable}

\section{Topic Selection Procedure for LDA}

\begin{figure}[h]
  \centering
  \includegraphics[width=\linewidth]{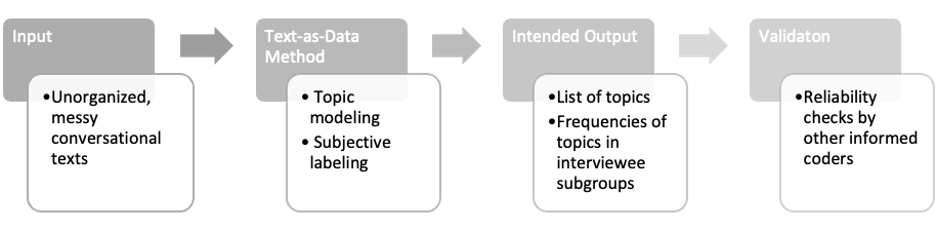}
\end{figure}

In this section, we will describe the steps that we took to adjust my topic models for effective and efficient subjective labeling. The labels produced in this section will reveal the lists of commonly appeared topics and anticipated topics. Then each document will be characterized by its proportions in different topics. Finally, those variables can be used in the inter- or intra-group analysis.

To incorporate covariates in topic models, we chose the Structural Topic Model (STM) package in R. we use sentences, including one and grouped sentences, as the unit of analysis. Short paragraphs will be easier to interpret, especially when the responses to a single question are discontinuous and unorganized. Moreover, sentence units contain grouped neighboring sentences that help mitigate meaning distortion in the interpretation. 

Starting with the cleaned and integrated data from the previous preprocessing steps, we pass the document-term matrix (generated by textProcessor()) through the searchK() function from stm packages. The function yields the diagnostic values by a number of manually input K-terms. We input a set of K equals (15, 17, 20, 23, 25, 27, 30, 33, 35) and obtain the output model statistics, including held-out likelihood, residuals, semantic coherence, and lower bound.

\begin{figure}[h]
  \centering
  \includegraphics[width=\linewidth]{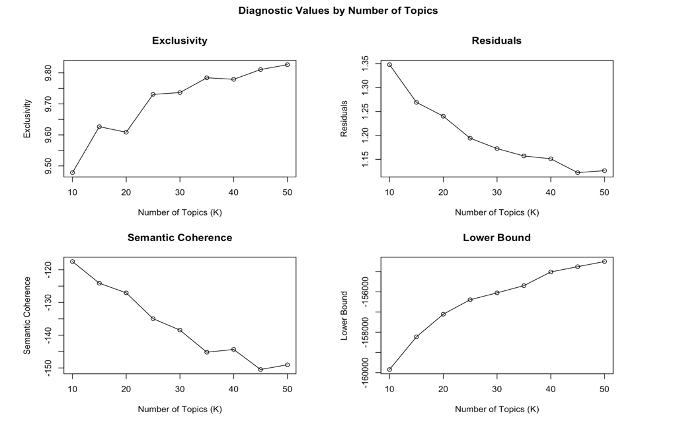}
  \caption{Diagnostic Values for a Number of Topics}
  \label{fig:lda_diag}
\end{figure}

\begin{figure}[h]
  \centering
  \includegraphics[width=0.7\linewidth]{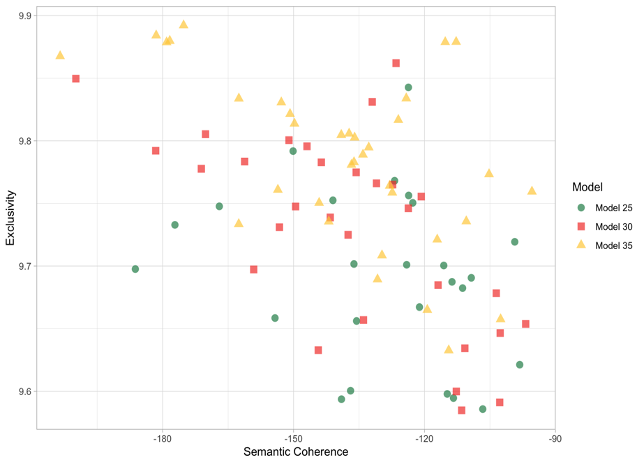}
  \caption{Exclusivity vs. Semantic Coherence}
  \label{fig:exclu_coh}
\end{figure}

Knowing that there is no single best model or “the” correct K terms, we evaluate models with different specifications following the previous studies that used the topic modeling method. Roberts et al. [2013]\cite{roberts2014structural} suggest selecting the model with lower residual value and higher held-out likelihood. Besides, a desirable model also needs to have high semantic coherence. The model statistics shown in Figure~\ref{fig:lda_diag} help narrow down the selection K ranges from 15 to 35. Models of more than or equal to 40 topics have good results for held-out likelihood and residual. But its semantic coherence is the lowest compared to models with fewer topics. Besides, considering the size of the data, a large K like 40 may result in a model that produces too narrow topics. In the second round, we run the diagnostic values for the numbers of topics within this new range and further narrow down the desirable numbers of topics to 24, 30, and 35 based on the same set of standards. Then, we compare exclusivity and semantic coherence for models using 24, 30, and 3 topics. As shown in Figure~\ref{fig:exclu_coh}, the model of 30 topics performs better in exclusivity, while the model of 25 topics performs better in semantic coherence. However, a smaller K tends to produce a high semantic coherence with probably too general topics \cite{aslett2020}, and it performs poorly on exclusivity.

After inspecting some sample topics for each model, we decided that the model of 30 topics captures a more comprehensive picture of the text corpus. 

\section{Additional Figures and Tables for Results Section}

\begin{figure}[h]
  \centering
  \includegraphics[height=0.45\textheight, keepaspectratio]{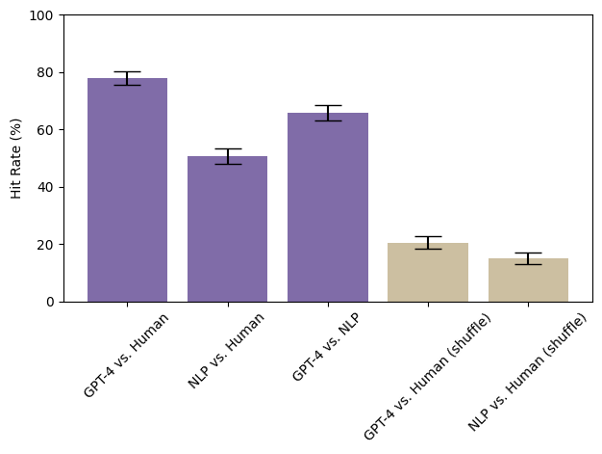}
  \includegraphics[height=0.45\textheight, keepaspectratio]{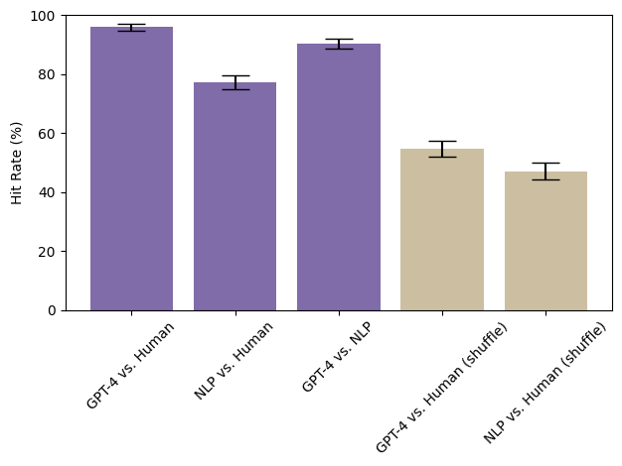}
  \caption{Comparison for Hit Rates and Shuffle Hit Rates with 95\% Confidence Intervals (a) Child codes (above) and (b) Parent codes (below)}
  \label{fig:c1}
\end{figure}

\begin{table}[h]
\centering
\caption{Robustness Tests}
\resizebox{\textwidth}{!}{
\begin{tabular}{lcccccc}
\toprule
& \multicolumn{3}{c}{Child codes} & \multicolumn{3}{c}{Parent codes} \\
\cmidrule(lr){2-4} \cmidrule(lr){5-7}
& Szymkiewicz-Simpson coefficients & S\o rensen-Dice coefficients & Jaccard similarity & Szymkiewicz-Simpson coefficients & S\o rensen-Dice coefficients & Jaccard similarity \\
\midrule
GPT-4 vs. Human & 0.63 & 0.42 & 0.30 & 0.87 & 0.62 & 0.49 \\
LDA vs. Human & 0.32 & 0.29 & 0.19 & 0.63 & 0.52 & 0.39 \\
\bottomrule
\end{tabular}
}
\label{tab:c1robust}
\end{table}

\begin{figure}[h]
  \centering
  \includegraphics[width=\linewidth]{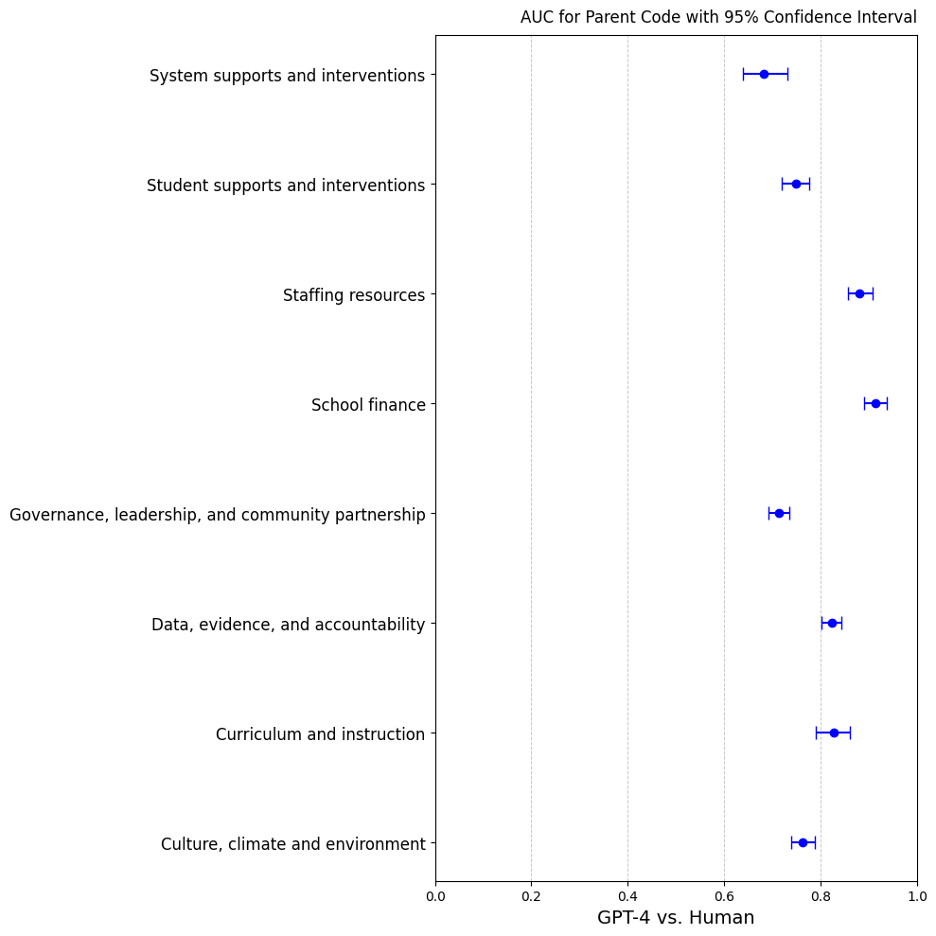}
  \caption{Parent-Code-Level Codewise AUCs with 95\% Confidence Intervals between GPT-4 and Human}
  \label{fig:c2}
\end{figure}

\begin{figure}[h]
  \centering
  \includegraphics[width=\linewidth]{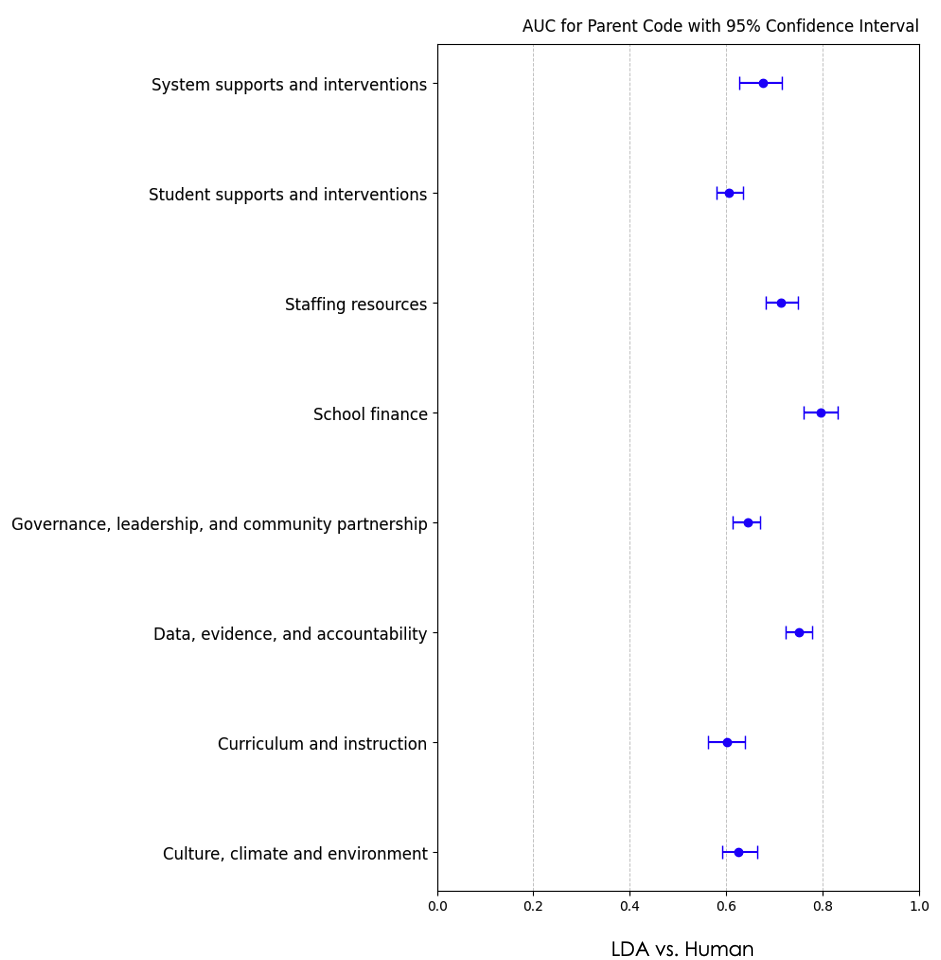}
  \caption{Parent-Code-Level Codewise AUCs with 95\% Confidence Intervals between LDA and Human}
  \label{fig:c3}
\end{figure}

\begin{figure}[h]
  \centering
  \includegraphics[width=\linewidth]{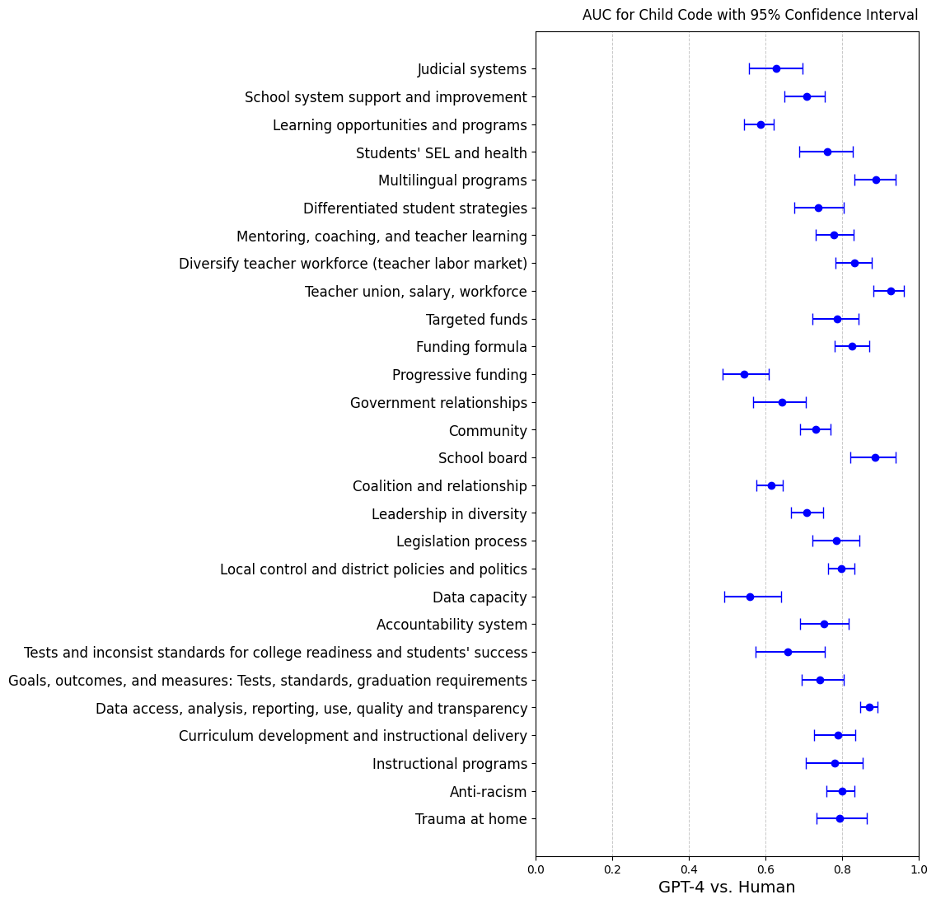}
  \caption{Child-Code-Level Codewise AUCs with 95\% Confidence Intervals Between GPT-4 and Human}
  \label{fig:c4}
\end{figure}

\begin{figure}[h]
  \centering
  \includegraphics[width=\linewidth]{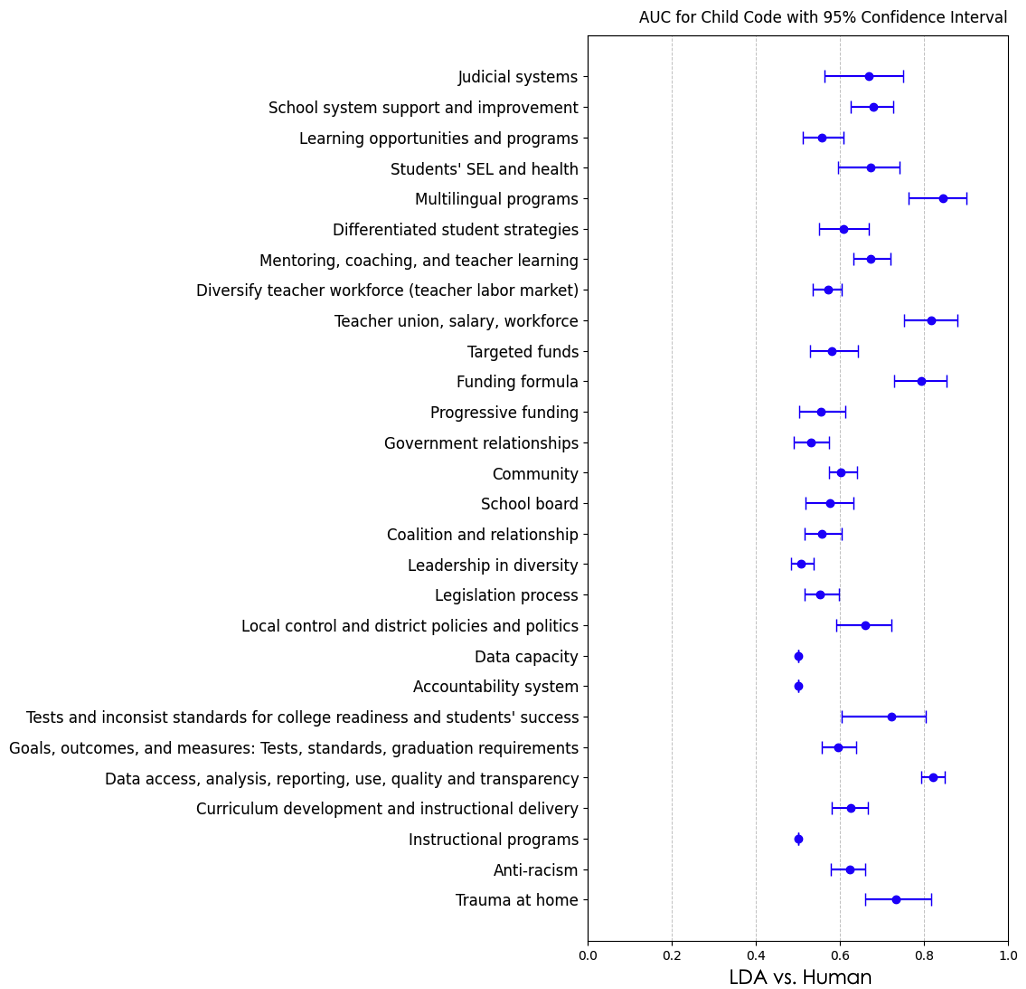}
  \caption{Child-Code-Level Codewise AUCs with 95\% Confidence Intervals Between LDA and Human}
  \label{fig:c5}
\end{figure}

\begin{figure}[h]
  \centering
  \includegraphics[width=\linewidth]{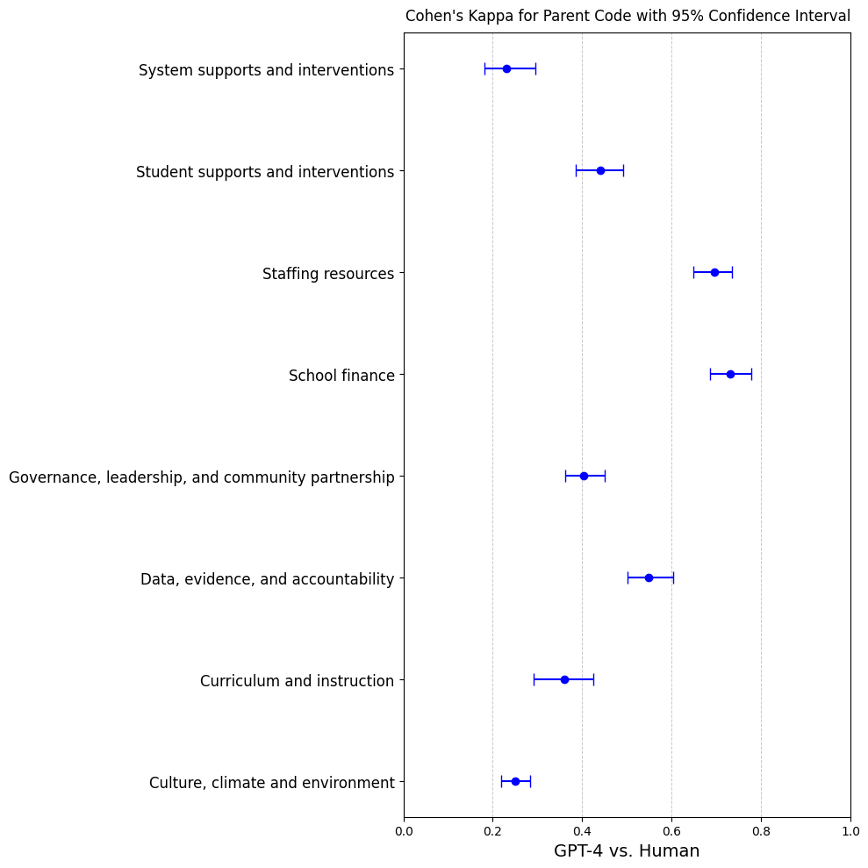}
  \caption{Parent-Code-Level Codewise Cohen’s $\kappa$ with 95\% Confidence Intervals between Gpt-4 and Human}
  \label{fig:c6}
\end{figure}

\begin{figure}[h]
  \centering
  \includegraphics[width=\linewidth]{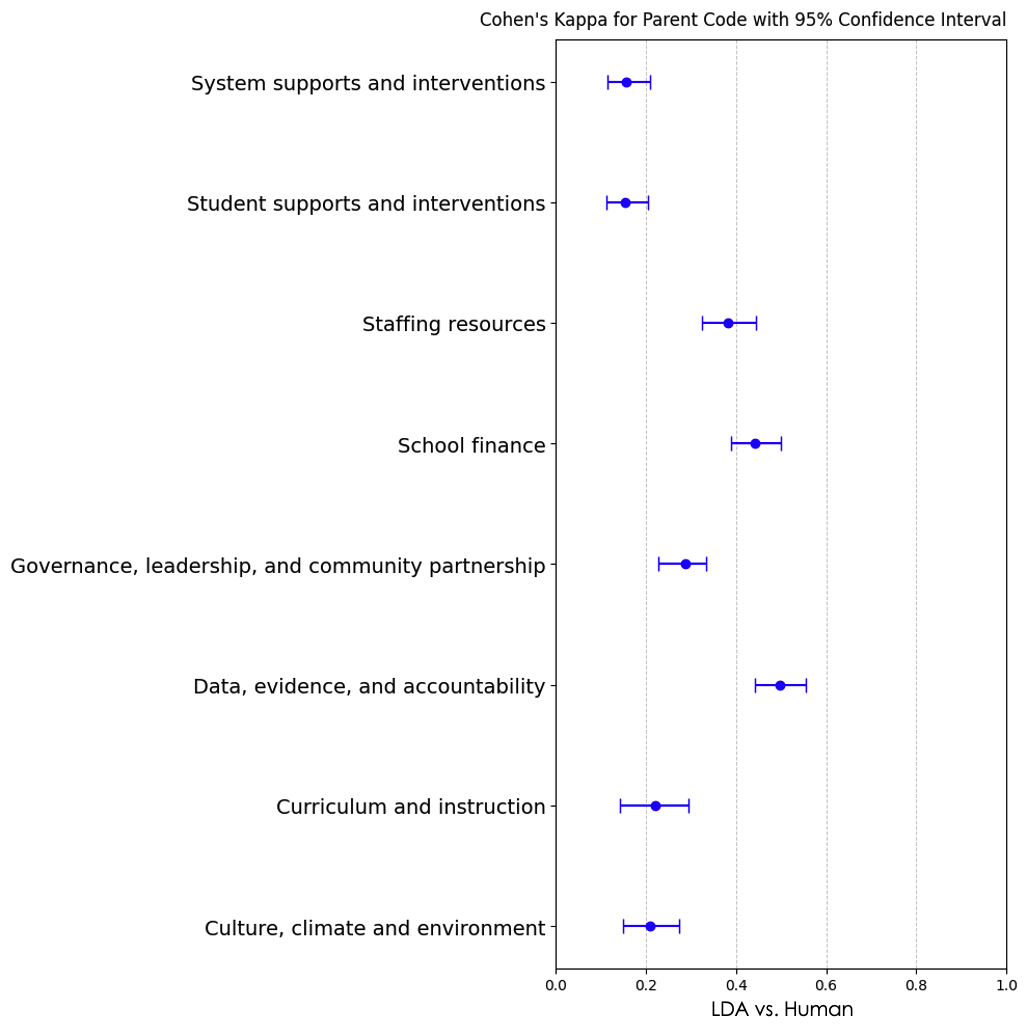}
  \caption{Parent-Code-Level Codewise Cohen’s $\kappa$ with 95\% Confidence Intervals Between LDA and Human}
  \label{fig:c7}
\end{figure}

\begin{figure}[h]
  \centering
  \includegraphics[width=\linewidth]{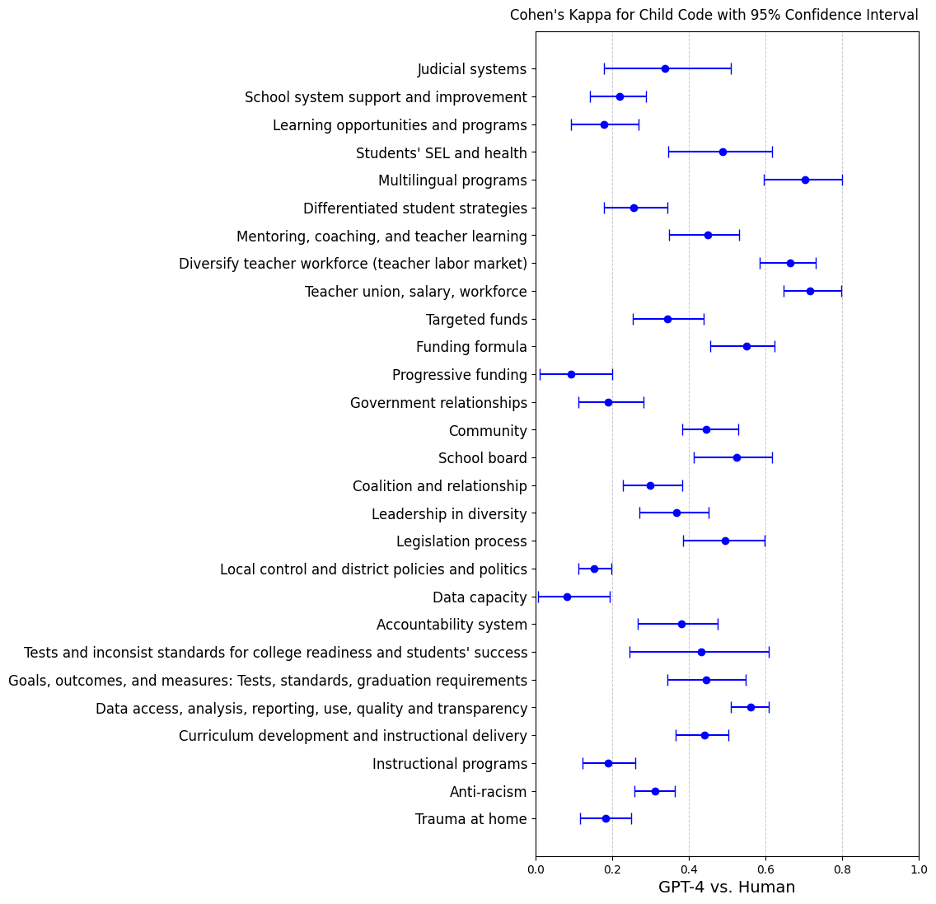}
  \caption{Child-Code-Level Codewise Cohen’s $\kappa$ with 95\% Confidence Intervals between GPT-4 and Human}
  \label{fig:c8}
\end{figure}

\begin{figure}[h]
  \centering
  \includegraphics[width=\linewidth]{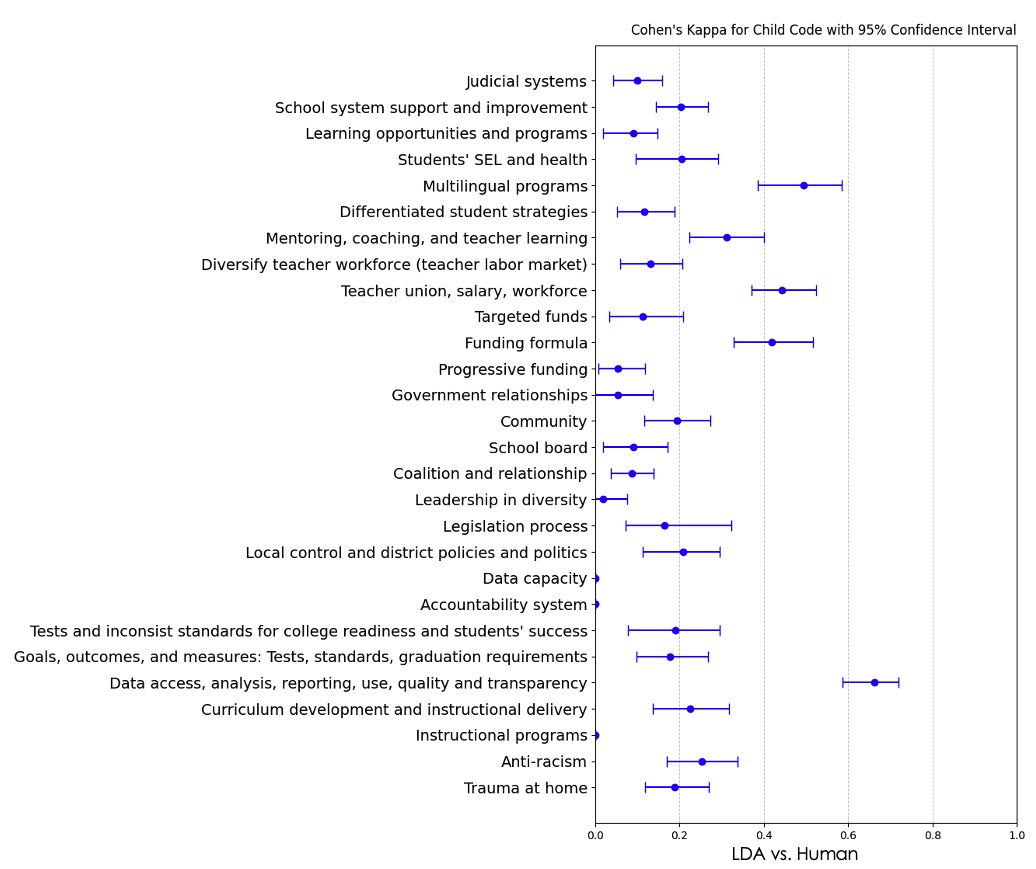}
  \caption{Child-Code-Level Codewise Cohen’s $\kappa$ with 95\% Confidence Intervals between LDA and Human}
  \label{fig:c9}
\end{figure}

\begin{figure}[h]
  \centering
  \includegraphics[width=0.9\linewidth]{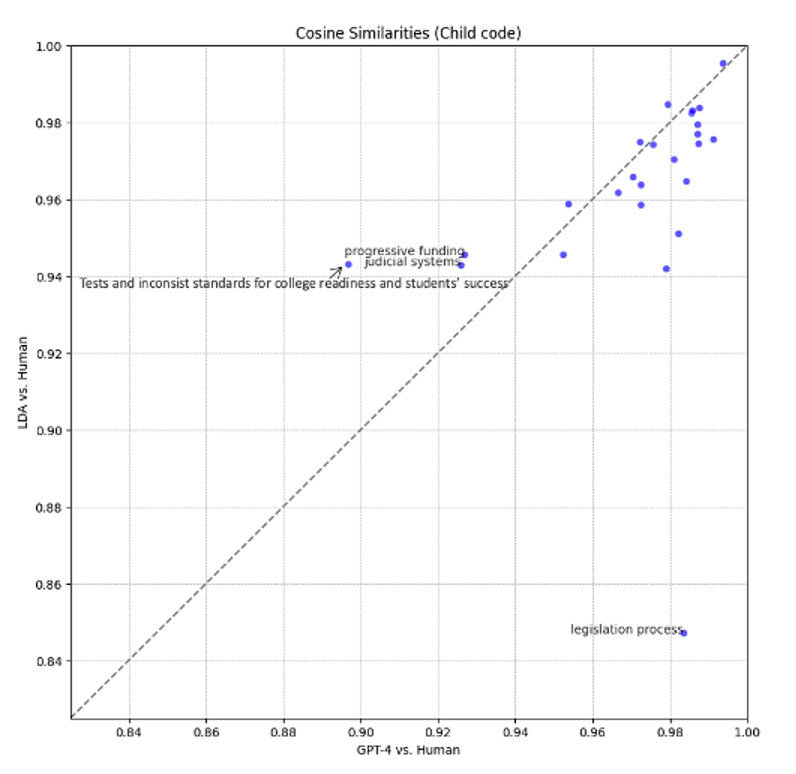}
  \caption{Child Code Cosine Similarity}
  \label{fig:c10}
\end{figure}

\begin{figure}[h]
  \centering
  \includegraphics[width=\linewidth]{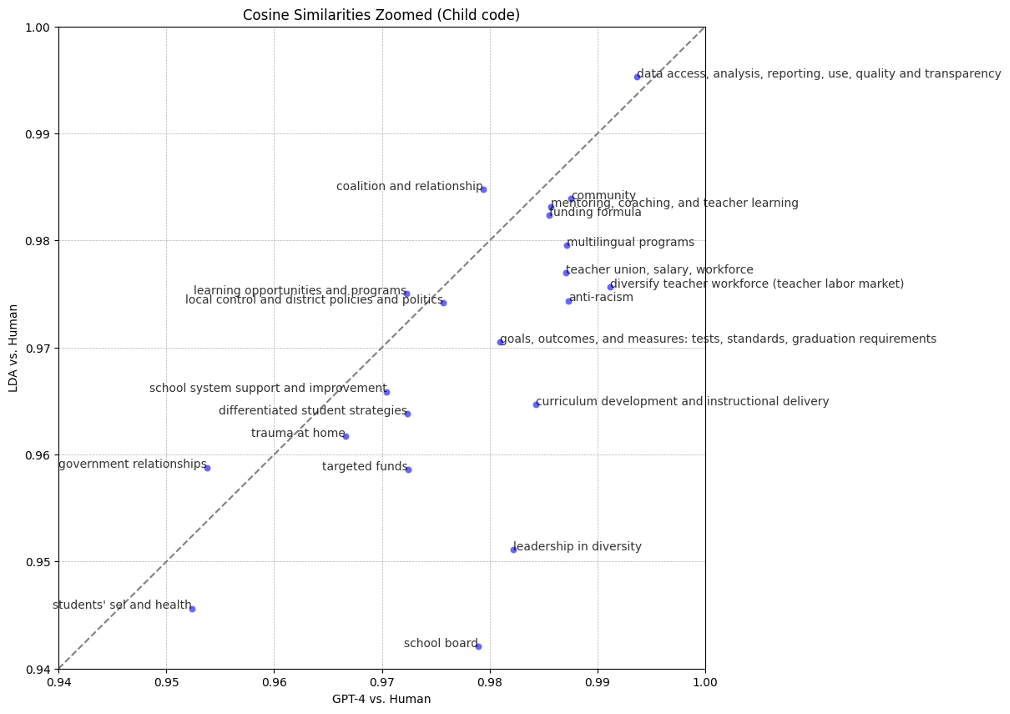}
  \caption{Zoomed-in Child Code Cosine Similarity}
  \label{fig:c11}
\end{figure}

\begin{figure}[h]
  \centering
  \includegraphics[width=\linewidth]{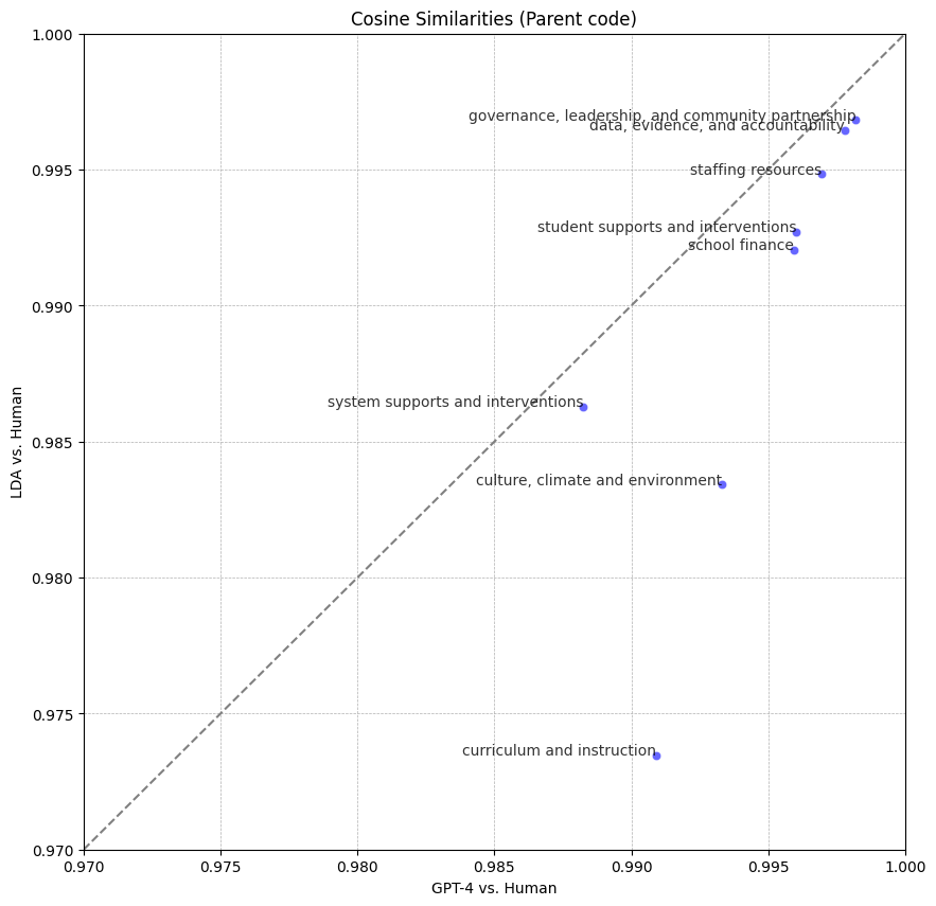}
  \caption{Parent Code Cosine Similarity}
  \label{fig:c12}
\end{figure}

\begin{figure}[h]
  \centering
  \includegraphics[width=\linewidth]{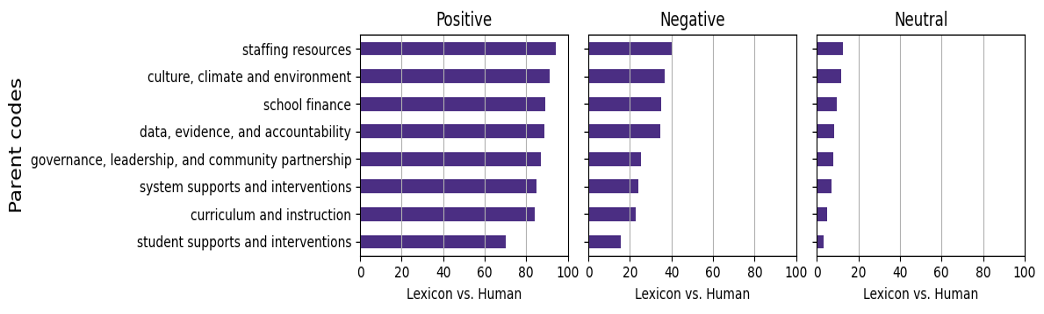}
  \includegraphics[width=\linewidth]{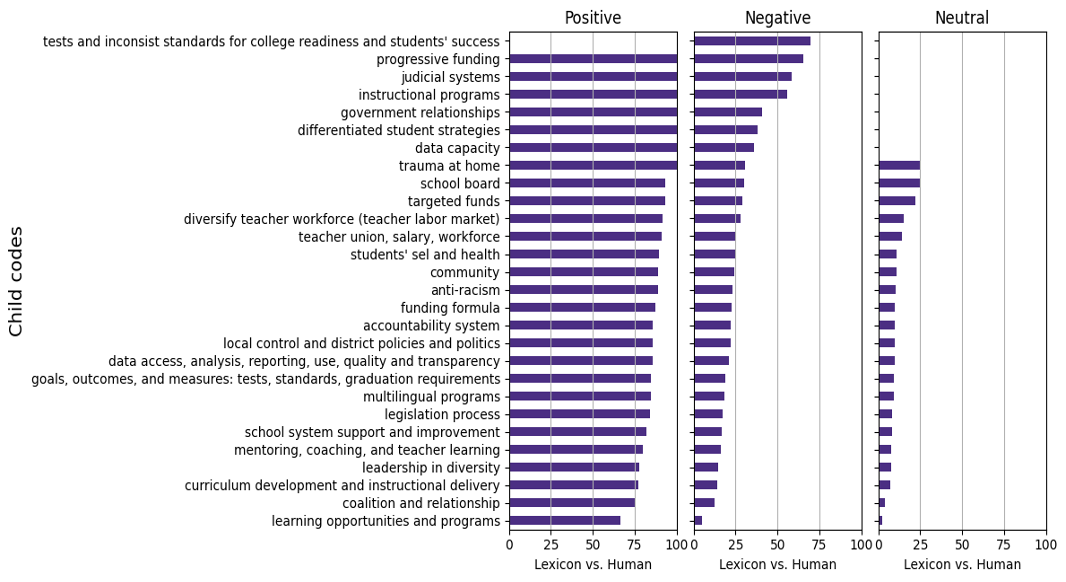}
  \caption{Percentage of Lexical-based Sentiment Annotation Agree with Human Annotation by Themes}
  \label{fig:c13}
\end{figure}

\FloatBarrier
\begin{longtable}{p{0.25\linewidth}p{0.15\linewidth}p{0.25\linewidth}p{0.15\linewidth}p{0.15\linewidth}}
\caption{GPT-4 Annotation (Original) Distribution Compared with Human Child Code Labels} \\
\toprule
Parent Codes & Parent Code Frequency (\%) & Child Codes & Child Code Frequency (\%) &	Human Labeled Child Code Frequency (\%) \\
\midrule
\endfirsthead
\multicolumn{5}{c}%
{\tablename\ \thetable\ -- \textit{Continued from previous page}} \\
\toprule
Parent Codes & Parent Code Frequency (\%) & Child Codes & Child Code Frequency (\%) &	Human Labeled Child Code Frequency (\%) \\
\midrule
\endhead
\hline \multicolumn{5}{r}{\textit{Continued on next page}} \\
\endfoot
\bottomrule
\endlastfoot

Data, evidence, and accountability & 9.51 & Data access, analysis, reporting, use, quality and transparency & 4.45 & 9.32 \\
Curriculum and instruction & 2.52 & Curriculum development and instructional delivery & 3.5 & 3.34 \\
Governance, leadership, and community partnership & 12.09 & Local control and district policies and politics & 3.5 & 2.64 \\
Governance, leadership, and community partnership & 12.09 & Community & 3.06 & 5.89 \\
Culture, climate and environment & 9.7 & Anti-racism & 2.52 & 5.06 \\
Governance, leadership, and community partnership & 12.09 & Leadership in diversity & 2.47 & 5.38 \\
Student supports and interventions & 3.01 & Differentiated student strategies & 2.41 & 2.5 \\
Governance, leadership, and community partnership & 12.09 & Government relationships & 2.22 & 2.18 \\
Staffing resources & 4.55 & Diversify teacher workforce (teacher labor market) & 2.09 & 4.78 \\
Staffing resources & 4.55 & Mentoring, coaching, and teacher learning & 2.09 & 4.08 \\
Student supports and interventions & 3.01 & Learning opportunities and programs & 2.03 & 4.31 \\
Governance, leadership, and community partnership & 12.09 & Legislation process & 2.03 & 2.78 \\
Data, evidence, and accountability & 9.51 & Accountability system & 1.95 & 2.55 \\
Staffing resources & 4.55 & Teacher union, salary, workforce & 1.6 & 2.74 \\
Curriculum and instruction & 2.52 & Instructional programs & 1.49 & 1.44 \\
School finance & 4.74 & Targeted funds & 1.44 & 2.18 \\
Student supports and interventions & 3.01 & Multilingual programs & 1.41 & 2.27 \\
System supports and interventions & 3.61 & School system support and improvement & 1.38 & 3.53 \\
Governance, leadership, and community partnership & 12.09 & School board & 1.25 & 1.67 \\
Student supports and interventions & 3.01 & Students' SEL and health & 1.17 & 1.99 \\
Governance, leadership, and community partnership & 12.09 & Coalition and relationship & 1.14 & 6.77 \\
School finance & 4.74 & Funding formula & 1.06 & 3.06 \\
Data, evidence, and accountability & 9.51 & Goals, outcomes, and measures: Tests, standards, graduation requirements & 1.06 & 3.57 \\
Culture, climate and environment & 9.7 & Trauma at home & 0.54 & 1.76 \\
School finance & 4.74 & Progressive funding & 0.43 & 1.58 \\
System supports and interventions & 3.61 & Judicial systems & 0.33 & 1.58 \\
Data, evidence, and accountability & 9.51 & Data capacity & 0.27 & 1.11 \\
Data, evidence, and accountability & 9.51 & Tests and inconsistent standards for college readiness and students' success & 0.27 & 1.16 \\

\end{longtable}

\end{document}